\documentclass[fleqn,usenatbib]{mnras}

\usepackage{newtxtext,newtxmath}
\usepackage{microtype}

\usepackage{graphicx}               % Including figure files
\usepackage[noend]{algpseudocode}
\usepackage{algorithm}
\usepackage{bm}

\hypersetup{colorlinks,urlcolor=magenta,linkcolor=red,citecolor=blue}
\newcommand{\dd}{{\,\mathop{\kern0pt\mathrm{d}}\!{}}}
\newcommand{\ppxf}{\textsc{ppxf}}
\newcommand{\norm}[1]{\left\lVert #1 \right\rVert}
\newcommand{\kms}{\ensuremath{\mathrm{km}\,s^{-1}}}		
\newcommand{\msun}{\ensuremath{\mathrm{M}_{\odot}}}
\newcommand{\lsun}{\ensuremath{\mathrm{L}_{\odot}}}

\def\equationautorefname~#1\null{equation~(#1)\null}

\title[\ppxf\ with spectra and photometry at $z\approx0.8$]{Full spectrum fitting with photometry in \ppxf: stellar population versus dynamical masses, non-parametric star formation history and metallicity for 3200 LEGA-C galaxies at redshift $z\approx0.8$}

\author[M.~Cappellari]{Michele Cappellari\thanks{E-mail: michele.cappellari@physics.ox.ac.uk}\\
	Sub-Department of Astrophysics, Department of Physics, University of Oxford, Denys Wilkinson Building, Keble Road, Oxford, OX1 3RH, UK}

\date{Accepted 2023 August 17. Received 2023 July 30; in original form 2022 September 1}

\pagerange{\pageref{firstpage}--\pageref{lastpage}} \pubyear{2023}

\begin{document}
\label{firstpage}
\maketitle

\begin{abstract}	
I introduce some improvements to the \ppxf\ method, which measures the stellar and gas kinematics, star formation history (SFH) and chemical composition of galaxies. I describe the new optimization algorithm that \ppxf\ uses and the changes I made to fit both spectra and photometry simultaneously. I apply the updated \ppxf\ method to a sample of 3200 galaxies at redshift $0.6<z<1$ (median $z=0.76$, stellar mass $M_\ast\ga3\times10^{10}$ M$_\odot$), using spectroscopy from the LEGA-C survey (DR3) and 28-bands photometry from two different sources. I compare the masses from new JAM dynamical models with the \ppxf\ stellar population $M_\ast$ and show the latter are more reliable than previous estimates. I use three different stellar population synthesis (SPS) models in \ppxf\ and both photometric sources. I confirm the main trend of the galaxies' global ages and metallicity $[M/H]$ with stellar velocity dispersion $\sigma_\ast$ (or central density), but I also find that $[M/H]$ depends on age at fixed $\sigma_\ast$. The SFHs reveal a sharp transition from star formation to quenching for galaxies with $\lg(\sigma_\ast/\kms)\ga2.3$ ($\sigma_\ast\ga200$ \kms), or average mass density within 1~kpc $\lg(\Sigma_1^{\rm JAM}/\mathrm{\msun kpc^{-2}})\ga9.9$ ($\Sigma_1^{\rm JAM}\ga7.9\times10^9\,\mathrm{\msun\ kpc^{-2}}$), or with $[M/H]\ga-0.1$, or with Sersic index $\lg n_{\rm Ser}\ga0.5$ ($n_{\rm Ser}\ga3.2$). However, the transition is smoother as a function of $M_\ast$. These results are consistent for two SPS models and both photometric sources, but they differ significantly from the third SPS model, which demonstrates the importance of comparing model assumptions. The \ppxf\ software is available from \url{https://pypi.org/project/ppxf/}.
\end{abstract}

\begin{keywords}
	galaxies: evolution -- galaxies: formation -- galaxies: high-redshift -- software: data analysis -- techniques: photometric -- techniques: spectroscopic
\end{keywords}

\section{Introduction}
\label{sec:intro}

The study of the stellar population of galaxies is an essential tool when trying to uncover how they have assembled. For this reason, a vast number of papers have tried to infer the galaxies' star formation history (SFH) and chemical composition from observations. The earliest results were based on simple galaxy colours as these were easier to obtain \citep[see e.g. the lectures by][]{Baade1963}. However, galaxy colours alone cannot strongly constrain both the galaxies' chemical composition and SFHs. Inferences from galaxy photometry alone are strongly affected for example by the age-metallicity \cite[e.g.][]{Worthey1994} as well as by the SFH-dust degeneracies \citep[e.g.][]{Silva1998,Devriendt1999,Pozzetti2000}. For this reason, most of our knowledge on both the star formation and chemical composition of galaxies has been obtained from the numerous absorption features in their spectra. 

\subsection{Full spectrum fitting of nearby galaxies}

Over the past two decades, libraries of high-resolution ($R\ga2000$) empirical stellar spectra were observed, which try to optimally sample all stages of stellar evolution. Prominent examples in the optical region include the STELIB \citep{LeBorgne2003}, ELODIE \citep{Prugniel2001}, MILES \citep{Sanchez-Blazquez2006,FalconBarroso2011miles} and recently the MaStar \citep{Yan2019} stellar libraries. An exception is the X-Shooter spectral library (XSL), which reaches $R\approx10000$ and extends up to 2.5 $\micron$ \citep{Chen2014,Verro2022a}.

Stellar population synthesis (SPS) models based on empirical stellar spectra have been developed that can produce synthetic galaxy spectra at high resolution. Examples of these kind of models are the \textsc{galaxev} \citep{Bruzual2003}, the Vazdekis \citep{Vazdekis2010,Vazdekis2015}, the \textsc{fsps}  \citep{Conroy2009,Conroy2010}, the Maraston \citep{Maraston2005,Maraston2011,Maraston2020} and most recently the XSL models \citep{Verro2022}. Most current SPS models complement empirical stellar spectra with fully synthetic ones like BaSeL \citep{Westera2002} or MARCS \citep{Gustafsson2008}. This allows one to cover stages of stellar evolution not well sampled by observations and also can extend the wavelength coverage, especially to the ultraviolet and infrared regions which are poorly covered by observations. Fully synthetic SPS were also developed like \textsc{bpass} \citep{Stanway2018,Byrne2022} or the MARCS version of the Maraston models.

Initially, large spectroscopic studies of the stellar population in nearby galaxies were based on line indices of specific absorption features, generally using the LICK system \cite[e.g.][]{Worthey1994a}, but the availability of good-quality high-resolution SPS models \citep[see review by][]{Conroy2013} motivated a shift to using full-spectrum fitting \citep[see review by][]{Walcher2011}. Various templates-fitting methods were developed for this task, like \ppxf\ \citep{Cappellari2004,Cappellari2017}, \textsc{starlight} \citep{CidFernandes2005}, \textsc{stecmap} \citep{Ocvirk2006}, \textsc{vespa} \citep{Tojeiro2007}, \textsc{fit3D} \citep{Sanchez2016,Lacerda2022} and \textsc{firefly} \citep{Wilkinson2017}. These methods and software were extensively used e.g. to analyse the millions of spectra produced by integral-field spectroscopic surveys in the local Universe like ATLAS$^{\rm 3D}$ \citep{Cappellari2011a}, CALIFA \citep{Sanchez2012}, SAMI \citep{Bryant2015} and MaNGA \citep{Bundy2015}. 

\subsection{Spectral fitting of high-$z$ galaxies}

At significant redshift (e.g. $z\ga1$) the cosmological surface brightness dimming \citep[e.g.][]{Hogg1999} makes good-quality spectra more difficult to obtain as the contribution of the sky background starts to dominate. \ppxf\ was used to measure the kinematic and stellar population of significant samples of galaxies at redshift $z\approx1$ \citep[e.g.][]{Shetty2015,Bezanson2018} but only of individual objects out to $z\approx2$ \citep[e.g.][]{vandeSande2013,Belli2014,Belli2017} and $z\approx3$ \citep[e.g.][]{Esdaile2021,Forrest2022}. Most large studies of distant galaxies still had to rely on photometry alone.

The most essential parameters one wants to extract from high-$z$ galaxies are their redshift and stellar mass \citep[e.g.][]{Muzzin2013,Weaver2022}. Various template-fitting codes were developed to measure masses and redshift from photometric observations in multiple bands (I ignore here methods based on machine learning; see \citealt{Salvato2019} for a review). These include \textsc{Hyperz} \citep{Bolzonella2000}, \textsc{bpz} \citep{Benitez2000}, \textsc{LePhare} \citep{Arnouts2002}, \textsc{zebra} \citep{Feldmann2006} and \textsc{eazy} \citep{Brammer2008}. These methods are conceptually similar to the template-based spectral fitting ones used for nearby galaxies, however, they all adopt a Bayesian approach, instead of a least-squares fitting one. This makes the codes simpler and allows for easy inclusion of priors on galaxy parameters or non-Gaussian uncertainties; e.g. one can assign a low probability to solutions where the galaxy has an unphysically large/small stellar mass. 

Building on the photometric-redshift and full-spectrum fitting approaches, new software was later developed to fit spectra together with the photometry, while still retaining the same Bayesian approach of photometric-redshift codes. Examples of these are \textsc{fast} \citep{Kriek2009}, \textsc{beagle} \citep{Chevallard2016}, \textsc{bagpipes} \citep{Carnall2018}, the code described by \citet{Mendel2020} and \textsc{prospector} \citep{Johnson2021}. 

Contrary to what is sometimes stated, \emph{both} least-squares, or maximum-likelihood, and Bayesian methods can return model posteriors, when needed. The former uses bootstrapping \citep[e.g.][]{Efron1994} or Monte Carlo approaches. In fact, bootstrapping can be seen as an efficient way to compute the Bayesian posterior, with non-informative priors \citep[e.g.][]{Rubin1981,Efron2011}. Although bootstrapping is less flexible than general Bayesian methods, in many realistic situations, the uncertainties of model parameters are dominated by data systematic and model assumptions (as I also find later) rather than the details of the adopted statistical approach or by adopted priors.

\subsection{This paper}

In this paper, I proceed differently than most existing methods. Instead of adopting the standard Bayesian approach to fit photometry and spectra, I present an extension of my \ppxf\ least-squares full-spectrum fitting method to simultaneously fit photometry. A key difference in this approach is that it can be a few orders of magnitude faster than Bayesian methods. Apart from algorithmic differences, my \ppxf\ approach to fitting photometry with spectra is analogue to the extension of the \textsc{starlight} least-squares full-spectrum fitting method \citep{LopezFernandez2016,Werle2019}. Lest-squares methods appear complementary to existing ones as the extra speed allows for extra flexibility in the treatment of the stellar population, as shown later.

I illustrate the characteristics of the approach by fitting the VIMOS spectra and COSMOS photometry \citep{Muzzin2013,Weaver2022} to study the joint SFH - metallicity distributions and the stellar population scaling relations of about 3200 galaxies from the LEGA-C survey \citep{vanderWel2021} in the redshift range $0.6<z<1$.

Readers interested in the \ppxf\ techniques should keep reading how to measure velocities in \autoref{sec:velocity} and the \ppxf\ updates in \autoref{sec:updates}. While those only interested in the scientific results should skip the next two sections and go directly to the description of the data in \autoref{sec:data} and results in \autoref{sec:results}. 
In this work, I adopt a standard cosmology with $H_0=70$ \kms Mpc$^{-1}$, $\Omega_m=0.3$ and $\Omega_\Lambda=0.7$.

\section{Measuring velocity and redshift}
\label{sec:velocity}

In \citet[sec.~2]{Cappellari2017} I reviewed general and important facts that one should know before using any full spectral fitting method and \ppxf\ in particular. Here I include only some updates, and I heavily refer the reader to my previous paper of this series to avoid duplicating material.

\subsection{From measured velocity to observed redshift}

The physical meaning of the velocity $V$ returned by \ppxf\ or any spectrum-fitting code is often a source of confusion. As discussed in \citet[sec.~2.3]{Cappellari2017}, the reason for this is that $V$ has no physical meaning. Even the recession velocity itself, for a distant galaxy, is an ill-defined concept with a debated interpretation \citep[e.g.][]{Bunn2009}. It should never be used for quantitative work. What is well-defined empirically is the redshift $z$ of a given spectrum:
\begin{equation}\label{eq:pixel_shift}
	1 + z \equiv \frac{\lambda_{\rm obsv}}{\lambda_{\rm emit}},
\end{equation}
where $\lambda_{\rm obsv}$ and $\lambda_{\rm emit}$ are the observed and rest-frame wavelength of a given spectral feature. The key formula that is needed to convert the $V_{\ppxf}$ returned by \ppxf\ into redshift is \citep[eq.~8]{Cappellari2017}
\begin{equation}\label{eq:vel_to_z}
	V_{\ppxf}\equiv c\, \ln(1+z),
\end{equation}
with $c$ the speed of light. This formula is exact by construction and it is the only one to use to attach a physical meaning to $V_{\ppxf}$.

\subsection{Separating peculiar velocities and cosmological redshift}

When observing spectra of distant galaxies from a single aperture, redshift is all one can measure. However, when obtaining spatially-resolved observations of galaxies e.g. using integral-field spectroscopy \citep[see review by][]{Cappellari2016} one needs to separate the cosmological redshift $z_{\rm cosm}$, which only contains information on the galaxy distance, from the peculiar velocity $V_{\rm pec}$. The latter is the one which satisfies e.g. Newton's law of gravitation in a reference system that moves with the galaxy barycentre. It is the velocity that has to be used to construct dynamical models of the galaxy.

In \citet[sec.~2.4]{Cappellari2017} I suggested using the standard way of separating peculiar and cosmological redshift. However, there is a simpler and formally even more accurate way. In fact, the conversion of $V_{\ppxf}$ into redshift is unnecessary \citep[see][]{Baldry2018}. One can directly obtain $V_{\rm pec}$ using the velocities returned by \ppxf\ as follows
\begin{equation}\label{eq:vel_pec}
	V_{\rm pec}(x,y) = V_{\ppxf}(x,y) - V_{\ppxf}({\rm bary}).
\end{equation}
Here $V_{\ppxf}(x,y)$ is the velocity returned by \ppxf\ at the location $(x,y)$ on the sky, $V_{\ppxf}({\rm bary})$ is the velocity returned by \ppxf\ for the galaxy (or cluster) barycentre and $V_{\rm pec}(x,y)$ is the peculiar velocity at location $(x,y)$. The latter is the only one with a clear physical meaning: it is the one to use in a dynamical model \citep[e.g.][]{Cappellari2008}, or to estimate the level of rotation in a galaxy \citep[e.g.][]{Emsellem2011}. Importantly, \autoref{eq:vel_pec} is always valid, regardless of whether the spectrum was de-redshifted to the rest-frame or not, before measuring $V_{\ppxf}$. Note that this formula only works because of the way \ppxf\ defines the relation between velocity and redshift in \autoref{eq:vel_to_z} and {\em cannot} be used with alternative definitions (e.g. $V\equiv c z$).

As an example of a practical application of these formulas, let's assume I am fitting a single spectrum of a high-$z$ galaxy for which I have an estimate of the redshift $z'$ (e.g. from photometry). It is generally convenient to de-redshift the spectrum by dividing each observed wavelength $\lambda_{\rm obs}$ to obtain an estimate of the rest-frame wavelength with
\begin{equation}
	\lambda'_{\rm rest}=\frac{\lambda_{\rm obs}}{(1 + z')}
\end{equation}
I then fit the spectrum with \ppxf\ to obtain $V_{\ppxf}$. If the initial guess $z'$ was perfect, I would obtain $V_{\ppxf}=0$, but in general I will measure $V_{\ppxf}\ne0$, with uncertainty $\Delta V_{\ppxf}$. An improved estimate of the galaxy redshift $z$ and its uncertainty $\Delta z$ can be obtained using \autoref{eq:vel_to_z} and \autoref{eq:vel_pec} as
\begin{subequations}\label{eq:z_example}
	\begin{align}
		&V' = c\, \ln(1+z'), \\
		&V_{\rm tot} = V_{\ppxf} + V'\\
		&1 + z = \exp\left(\frac{V_{\rm tot}}{c}\right) = (1 + z')\exp\left(\frac{V_\ppxf}{c}\right) \\
		&\frac{\Delta z}{1+z} \approx \Delta\ln(1+z) \approx\frac{\Delta V_{\ppxf}}{c}.
	\end{align}
\end{subequations}
Identical results are obtained without first bringing the spectrum to the restframe and setting $z'=0$ in \autoref{eq:z_example}. But one should remember to adjust the instrumental resolution as described in sec.~2.4 of \citet{Cappellari2017}.

\section{Updates to the \ppxf\ package}
\label{sec:updates}

I gave a detailed overview of the \ppxf\ method in \citet[sec.~3]{Cappellari2017}. I will not repeat that summary here, but instead, I will refer the reader to specific sections of that paper, while trying to keep a consistent notation. However, I substantially evolved \ppxf\ since then, driven by the needs of my research and requests from colleagues. I only describe here the \ppxf\ features that have changed.  The description corresponds to the current version 8.2 of the public \ppxf\ Python package\footnote{Available from \url{https://pypi.org/project/ppxf/}}.

\subsection{Well-sampled variable-$\sigma$ convolution}

As discussed in \citet[sec.~2.2]{Cappellari2017}, when fitting stellar templates to galaxy spectra, one generally needs to first match their resolution to that of the galaxy observations by convolving the templates with a Gaussian with dispersion $\sigma$ that varies with wavelength $\lambda$. In a previous version of \ppxf, I implemented this step as a direct summation of the spectrum, weighted by the Gaussian centred on every pixel (function \texttt{ppxf\_util.gaussian\_filter1d} in \ppxf). Given that the Gaussian is typically nonzero only over $n_{\rm gau}\sim10$ pixels, while the templates have $n_{\rm pix}\ga1000$ spectral pixels, I limited the summation only to the nonzero pixels of the Gaussian kernel. In this way, the computation time of this summation scales as $t\propto n_{\rm pix}\times n_{\rm gau}$, which is comparable to that $t\propto n_{\rm pix}\times\ln(n_{\rm pix})$ achievable using Fourier convolution using the Fast Fourier Transform (FFT, \citealt{Cooley1965fft}).

A limitation of performing a convolution as a summation is that, when the $\sigma$ of the Gaussian becomes comparable to the size of the sampled pixels, the convolution suffers from the same undersampling problems that motivated the use of an analytic Fourier Transform (FT) for the convolution in \ppxf\ discussed in \citet{Cappellari2017}. For this reason, I now implemented an alternative procedure \texttt{ppxf\_util.varsmooth} which performs the variable-$\sigma$ convolution using the FFT and uses the same approach of \ppxf\ of using an analytic Fourier Transform of the kernel, to avoid undersampling issues.

A natural idea to use the FFT for the convolution of a vector of values with a kernel with variable scale is to stretch the coordinate with the inverse of the scale, via interpolation, in such a way that the kernel has the same scale in the new coordinates. This idea was discussed e.g. in 2014 on StackOverflow\footnote{\url{https://stackoverflow.com/a/24186800}} and implemented in 2016 by Janez Kos on GitHub\footnote{\url{https://github.com/sheliak/varconvolve}} in the procedure \texttt{varconvolve}. It was also discussed in \citet{Johnson2021}. The interpolation approach is also central for algorithms for the Non-Uniform FFT \citep[NUFFT, e.g.][]{Greengard2004}. Here I combine the interpolation approach with the use of an analytic Fourier Transform as in \citet{Cappellari2017} to produce \autoref{alg:varsmooth}. The algorithm can also be used with a non-Gaussian kernel \citep[e.g.][eq.~38]{Cappellari2017} as long as only its scale changes with wavelength.

\begin{algorithm}
	\caption{\texttt{varsmooth}: well-sampled variable-$\sigma$ convolution}\label{alg:varsmooth}
	\begin{algorithmic}
		\State Given $\mathbf{x}, \mathbf{y}, \bm{\sigma}_x, \mathbf{x}_{\rm out}$ 
		\Comment ${\rm len}(\mathbf{x}) = {\rm len}(\mathbf{y}) = {\rm len}(\bm{\sigma}_x) = p$
		\State $\bm{\sigma} = \bm{\sigma}_x/\nabla\mathbf{x}$  \Comment{Convert $\bm{\sigma}_x$ into pixels\footnotemark}
		\State $\sigma_{\rm max} = \max(\bm{\sigma})\times m$ \Comment{Optional oversampling $m\ge1$}
		\State $\mathbf{s}={\rm cumsum}(\sigma_{\rm max}/\bm{\sigma})$  \Comment{Cumulative sum}
		\State $n = {\rm ceil}(s_p - s_1)$ \Comment{Ensure oversampling}
		\State $\mathbf{x}_{\rm new} = {\rm subdivide}(s_1, s_p, n)$  \Comment{$n$ equispaced values}
		\State $\mathbf{y}_{\rm new} = {\rm interpolate}(\mathbf{s}, \mathbf{y})(\mathbf{x}_{\rm new})$
		\State $\mathcal{F}({\rm gauss}) = \exp(-\sigma_{\rm max}^2\bm{\omega}^2/2)/\sqrt{2\pi}$  \Comment{Gaussian Analytic FT}
		\State $\mathbf{y}_{\rm conv} = \mathcal{F}^{-1}[\mathcal{F}(\mathbf{y}_{\rm new})\mathcal{F}({\rm gauss})]$
		\Comment{Convolution theorem \& FFT}
		\If{$\mathbf{x}_{\rm out}$ is given}
		\State $\mathbf{s} = {\rm interpolate}(\mathbf{x}, \mathbf{s})(\mathbf{x}_{\rm out})$
		\EndIf
		\State $\mathbf{y}_{\rm conv} = 
		{\rm interpolate}(\mathbf{x}_{\rm new}, \mathbf{y}_{\rm conv})(\mathbf{s})$
	\end{algorithmic}
\end{algorithm}
\footnotetext{I use a centred finite-difference approximation of the gradient $\nabla x$ instead of the spectral pixels size $\Delta x$, which is generally not provided. The two quantities coincide in uniformly-sampled regions of the spectrum.}

This algorithm works well in practice, but one should be aware of its theoretical limitations. In fact, most interpolation methods can be described as a convolution with a specific kernel \citep[e.g.][]{Getreuer2011} and one may think it would be better to remove the effect of this extra convolution as done e.g. in the NUFFT methods. However, this situation is different as the spectra have noise and one would have to perform the interpolation in a Bayesian framework \citep[e.g.][]{MacKay1992,MacKay2003}. One should also consider that the spectra to fit generally already include additional interpolation and resampling, which would have to be modelled for rigorous results. But all this is unlikely to affect scientific results, and for this reason, is beyond the scope of this paper.

\subsection{\textsc{CapFit} nonlinear least-squares with linear constraints}
\label{sec:capfit}

\subsubsection{The problem}

When fitting the kinematics of multiple kinematics components with \ppxf, both for the stellar and gas emission components, it is often useful to be able to set constraints on some parameters as a function of other parameters. For example, when looking for spectra containing both narrow and broad gas emission lines to study Active Galactic Nuclei \cite[AGN, e.g.][]{Oh2015,Fu2023}, to avoid the degeneracy of fitting two similar lines, one may want to constrain the dispersion of the broad emission component, if present, to be significantly larger than that of the narrow one $\sigma_{\rm broad}>\sigma_{\rm narrow}+\Delta\sigma$, or as a fractional difference $\sigma_{\rm broad}>f\times\sigma_{\rm narrow}$. Or one may want to constrain the velocity of a possible broad emission component to differ less than a certain value from that of the narrow one $|V_{\rm broad}-V_{\rm narrow}|<\Delta V$. Efficiently setting this kind of constraint requires solving a constrained non-linear optimization problem. 

For maximum computational efficiency and accuracy, one should exploit the special problem that \ppxf\ has to solve \citep[sec.~3.4]{Cappellari2017}. In particular, the function to minimize $f(\mathbf{x})$ is a sum of squares and the typical constraints are linear. The problem to solve can be expressed as
\begin{align}\label{eq:linear_system}
	{\rm minimize}\quad & f(\mathbf{x})=\tfrac{1}{2}\norm{\mathbf{r(\mathbf{x})}}^2\nonumber\\
	{\rm subject\; to}\quad & \mathbf{A}_{\rm eq}\cdot\mathbf{x}=\mathbf{b}_{\rm eq}\\
	& \mathbf{A}_{\rm ineq}\cdot\mathbf{x}\le\mathbf{b}_{\rm ineq}	\nonumber,
\end{align}
where $\mathbf{r(\mathbf{x})}$ are the residual from the fit, and $\mathbf{x}$ are the nonlinear parameters, like the kinematics of different components.
I have searched extensively for specialized software or an algorithm that I could easily use in \ppxf\ to efficiently solve this specific problem but did not find any. For this reason, I developed my own.

One of the most effective ways of solving nonlinear problems with general constraints is the sequential quadratic programming (SQP) method, where at every iteration the algorithm solves a constrained quadratic problem that approximates the function at the current location \citep[e.g.][chap.~18]{Nocedal2006}. 

In least-squares problems, one can approximate the function near the current point $\mathbf{x}_k$ as a second-order Taylor series, with $\mathbf{p}=\mathbf{x}-\mathbf{x}_k$, as follows \citep[e.g.][sec.~10.2]{Nocedal2006}
\begin{subequations}\label{eq:function_approx}
	\begin{align}
		f(\mathbf{x}) \approx\,& \tfrac{1}{2}\norm{\mathbf{J}_k\cdot\mathbf{p}+\mathbf{r}_k}^2\\
		=\,& \tfrac{1}{2}\norm{\mathbf{r}_k}^2 + \mathbf{p}\cdot(\mathbf{J}_k^T\cdot\mathbf{r}_k) + \tfrac{1}{2}\mathbf{p}\cdot(\mathbf{J}_k^T\cdot\mathbf{J}_k)\cdot\mathbf{p}\label{eq:quadratic}\\
		\approx\,& f(\mathbf{x}_k) + \mathbf{p}\cdot\nabla f(\mathbf{x}_k) + \tfrac{1}{2}\mathbf{p}\cdot\nabla^2 f(\mathbf{x}_k)\cdot\mathbf{p},
	\end{align}
\end{subequations}
where $\mathbf{J}_k$ is the Jacobian, which can be computed by finite differences, $\nabla f(\mathbf{x}_k)=\mathbf{J}^T_k\cdot\mathbf{r}_k$ is the gradient and $\nabla^2 f(\mathbf{x}_k)=\mathbf{J}_k^T\cdot\mathbf{J}_k$ is the \emph{quasi-Newton} approximation of the Hessian matrix, whose full form is the following, but I ignored the second term \citep[e.g.][eq.~10.5]{Nocedal2006}
\begin{equation}\label{eq:hessian}
	\nabla^2 f(\mathbf{x})=\mathbf{J}^T\cdot\mathbf{J}+\sum_{j=1}^m r_j \nabla^2 r_j.
\end{equation} 

It is a characteristic of least-squares problems that one can approximate the Hessian ``for free'' using $\mathbf{J}$ and the reason it is important to adopt specialized methods for their solution. In the case of \ppxf, the Hessian approximation is always especially good, even far from the solution, because the algorithm separates the linear and nonlinear optimizations \citep[sec.~3.3--3.4]{Cappellari2017} and ensures that $\sum_j r_j=0$ at every step. This tends to cancel out the second term in the Hessian of \autoref{eq:hessian}.

In essence, a specialized SQP algorithm to solve \autoref{eq:linear_system} would consist of solving a sequence of quadratic sub-problems as follows
\begin{align}\label{eq:sub_problem}
	{\rm minimize}\quad & g(\mathbf{p})=\norm{\mathbf{J}_k\cdot\mathbf{p}+\mathbf{r}_k}^2\nonumber\\
	{\rm subject\; to}\quad & \mathbf{A}_{\rm eq}\cdot\mathbf{x}=\mathbf{b}_{\rm eq}\\
	& \mathbf{A}_{\rm ineq}\cdot\mathbf{x}\le\mathbf{b}_{\rm ineq}	\nonumber.
\end{align}
Algorithms for this type of problem are discussed e.g. by \citet[sec.~11.2]{Fletcher1987} or \citet[chapter~5]{Gill1981}. I implemented those ideas into a trust-region algorithm \citep[chap.~4]{Nocedal2006} but I discovered that the approach is not sufficiently robust for my rather special situation. 

The difficulty of the optimization problem I have to solve consists of the fact that the Jacobian can sometimes be completely degenerate. A common situation where this happens is when \ppxf\ is fitting for the kinematics of emission lines or multiple stellar kinematic components. In this situation, the weights associated with a given emission line or stellar component may become exactly zero, because the line or component is simply not present in a certain galaxy spectrum. In other cases, the signal-to-noise ratio $S/N$ may be too low to give any constraints to some parameters. In these cases, the gradient (column of $\mathbf{J}$) with respect to the parameters describing the kinematics of the missing component will be zero.

I tried using the Singular Value Decomposition \citep[SVD, e.g.][sec.~15.4.2]{Press2007} during the iterations required to solve the quadratic sub-problem. But after extensive testing, e.g. during the development of the MaNGA Data Analysis Pipeline \citep{Westfall2019}, I was unable to find a robust criterion to decide which singular values must be edited and find the effective rank of my Jacobian.

\subsubsection{The solution}

To solve {\em unconstrained} least-squares optimization problems one of the most widely used techniques is the  Levenberg-Marquardt (LM) method \citep{Levenberg1944,Marquardt1963} and its state-of-the-art implementation in \textsc{minpack} \citep{More1978,More1980minpack}. The success of the LM method comes from the fact that the method penalizes the $\mathbf{J}$ matrix adaptively defining the quadratic sub-problem, in such a way that it always prevents degeneracy. This also makes the LM method a robust trust-region algorithm, as discussed in \citet[sec.~5.2]{Fletcher1987} or \citet[sec.~10.3]{Nocedal2006}. \citet[sec.~15.5.2]{Press2007} provides a less technical description.

In a previous version ($<$6.5) of \ppxf, I used the LM algorithm, as modified in the \textsc{mpfit} implementation \citep{Markwardt2009}, which included a very useful but non-optimal treatment of box constraints (i.e. upper/lower limits on the parameters). Comparable box-constrained least-squares methods exist in Scipy \citep{Scipy2020} as implemented in the trust-region reflective algorithm (\texttt{method=`trf`; \citealt{branch1999subspace}}) and the dogleg algorithm (\texttt{method=`dogbox`}, \citealt{voglis2004dogleg}, \citealt[chapter~4]{Nocedal2006}) in \href{https://docs.scipy.org/doc/scipy/reference/generated/scipy.optimize.least_squares.html}{\texttt{scipy.optimize.least\_squares}}. These methods are also available in \ppxf\ but cannot support linear constraints.

After extensive experimentation with real-world cases, I implemented a novel hybrid between the SQP and LM methods, specialized for the nonlinear least-squares with linear constraints (both equality and inequality). The algorithm consists of a trust-region quasi-Newton SQP method, with linear constraints, in which the matrix defining the quadratic sub-problem is penalized to avoid the risk of degeneracy, as in the LM method. I achieve this by replacing the quadratic sub-problem of \autoref{eq:sub_problem} with the following \citep[see][eq.~10.41]{Nocedal2006}
\begin{align}\label{eq:sub_problem_regul}
	{\rm minimize} \quad & g(\mathbf{p})=\norm{\left(
		\begin{array}{c}
			\mathbf{J}_k\\
			\sqrt{\lambda_k}\, \mathbf{D}_k
		\end{array}
		\right)
		\cdot\mathbf{p} +
		\left(
		\begin{array}{c}
			\mathbf{r}_k\\
			0
		\end{array}
		\right)}^2\nonumber\\
	{\rm subject\; to}\quad & \mathbf{A}_{\rm eq}\cdot\mathbf{x}=\mathbf{b}_{\rm eq}\\
	& \mathbf{A}_{\rm ineq}\cdot\mathbf{x}\le\mathbf{b}_{\rm ineq}	\nonumber,
\end{align}
where $\mathbf{D_k}$ is a diagonal matrix, which makes the problem scale invariant. By default, the diagonal elements of $\mathbf{D_k}$ are initialized with the norm $\norm{\cdot}$ of the columns of $\mathbf{J}$ and are updated during the iterations as suggested in \citet[eq.~6.3]{More1978}.
Close to the solution, when the quadratic model provides a good approximation of $f(\mathbf{x})$, then $\lambda_{k}$ becomes small and \autoref{eq:sub_problem_regul} approximates \autoref{eq:sub_problem}. In this limit, the method behaves as an SQP method. When the quadratic approximation is inaccurate, $\lambda_{k}$ becomes large and the method behaves as a trust-region LM method.

My resulting algorithm is rather simple because I did not worry about the efficiency of the solution of the quadratic programming sub-problem. The latter generally dominates the complexity of other state-of-the-art algorithms, which devise approximated matrix updates to save computation time \citep[e.g. see the description of LM in][]{More1978}.
I also did not try to deal with large-scale problems and sparse matrices which also increase complexity and require specialized methods \citep[e.g][]{Gill2005}. Instead, I focused on the fitting of rather small nonlinear problems ($n\la50$ variables) in which computing the function $f(\mathbf{x})$ involves creating a complex model, as in \ppxf. In this rather common situation, the time to solve the small quadratic programming sub-problem becomes negligible compared to that of evaluating $f(\mathbf{x})$. The solution is given in \autoref{alg:CapFit}, which I implemented in the \texttt{capfit} procedure in the \ppxf\ package.

\begin{algorithm}
	\caption{\textsc{CapFit}: linearly-constrained nonlinear least-squares}\label{alg:CapFit}
	\begin{algorithmic}
		\State Given $\mathbf{x}_1$, $f$, $\lambda>0$, $\eta\in[0,1/4)$
		\State $\mathbf{r}_1 = f(\mathbf{x}_1)$
		\State $\mathbf{J} = \mathbf{J}(\mathbf{x}_1)$
		\State Compute the scaling matrix $\mathbf{D}$ \Comment{See text}
		\Loop
		\State Obtain $\mathbf{p}$ as solution of \autoref{eq:sub_problem_regul} 
		with $\mathbf{J},\mathbf{D},\mathbf{r}_1,\lambda$
		\State $\mathbf{x}_2 = \mathbf{x}_1 + \mathbf{p}$
		\State $\mathbf{r}_2 = f(\mathbf{x}_2)$
		\State $r_{\rm act} = \chi^2(\mathbf{r}_1) - \chi^2(\mathbf{r}_2)$  \Comment{Using $\chi^2(\mathbf{x}) \equiv \mathbf{x}\cdot\mathbf{x}$}
		\State $r_{\rm pre} = \chi^2(\mathbf{r}_1) - \chi^2(\mathbf{J}\cdot\mathbf{p} + \mathbf{r}_1)$
		\State $\rho = r_{\rm act}/r_{\rm pre}$  \Comment{Actual vs predicted reduction}
		\If{convergence test is satisfied}
		\State \textbf{stop} with solution $\mathbf{x}_2$
		\EndIf
		\If{$\rho<1/4$} 
		\State $\lambda = 4\lambda$
		\ElsIf{$\rho>3/4$}
		\State $\lambda = \lambda/2$	
		\EndIf 
		\If{$\rho>\eta$} \Comment{Successful step: move on}
		\State $\mathbf{J}=\mathbf{J}(\mathbf{x}_2)$
		\State Adjust the scaling matrix $\mathbf{D}$ \Comment{See text}
		\State $\mathbf{x}_1, \mathbf{r}_1 = \mathbf{x}_2, \mathbf{r}_2$ 
		\EndIf
		\EndLoop
	\end{algorithmic}
\end{algorithm}

Except for the fact that the quadratic sub-problem is penalized and linearly-constrained, the algorithm uses the standard trust-region framework (e.g. \citealt[algorithm~4.1]{Nocedal2006}, or \citealt[algorithm 5.2.7]{Fletcher1987}). For the convergence criteria, I follow the description in \citet[sec.~2.3]{More1980minpack}.

\subsubsection{Solving the quadratic sub-problem}

I implemented two procedures to solve the quadratic programming sub-problem of \autoref{eq:sub_problem_regul}. In both cases, I avoid explicitly constructing the Hessian $\mathbf{J}_k^T\cdot\mathbf{J}_k$ as this would degrade the conditioning of the system. The first procedure (\texttt{lsq\_box}) is specialized for the common situation where only box constraints are present. It solves $\min \norm{\mathbf{A}\cdot\mathbf{x} - \mathbf{b}}^2$ with $\mathbf{lb}<\mathbf{x}<\mathbf{ub}$. For this, I use the active-set method adopted in the non-negative least-squares (\textsc{nnsl}) method \citep[algorithm~23.10]{Lawson1995}, which was generalized for box constraints with the Bounded-Variables Least-Squares (\textsc{bvls}) procedure in the same book and in \citet{Stark1995}. My implementation closely follows \citet{Lawson1995} except for the important fact that (i) I allow for a starting guess and (ii) I include an initialization step (\autoref{alg:lsq_box}) which generalizes to the box-constrained case the initialization loop in the \textsc{fastnnls} code\footnote{Available from \url{https://ucphchemometrics.com/}} by \citet{Andersson2000}. In realistic \ppxf\ problems, using my new \texttt{lsq\_box} with hundreds of spectral templates produced a typical speedup of a factor four compared to using the \href{https://docs.scipy.org/doc/scipy/reference/generated/scipy.optimize.nnls.html}{\texttt{scipy.optimize.nnls}}, which is a wrapper to the \citet{Lawson1995} Fortran code. The procedure \href{https://docs.scipy.org/doc/scipy/reference/generated/scipy.optimize.lsq_linear.html}{\texttt{scipy.optimize.lsq\_linear}} currently also does not support passing a starting guess.

\begin{algorithm}
	\caption{Initializiation \texttt{lsq\_box} box-constrained least-squares}\label{alg:lsq_box}
	\begin{algorithmic}
		\State Given $\mathbf{x}$, $\mathbf{A}$, $\mathbf{b}$, lower $\mathbf{lb}$ and upper $\mathbf{ub}$ bounds, $\mathcal{B}\ne\emptyset$
		\While{$\mathcal{B}\ne\emptyset$}
		\State $\mathcal{B}=\{\mathbf{x}\, |\, (\mathbf{x}<\mathbf{lb}) \lor (\mathbf{ub}<\mathbf{x})\}$ 
		\State Set all $\mathbf{x}\in\mathcal{B}$ to the nearest bound
		\State $\mathcal{F}=\{\mathbf{x}\, |\, \mathbf{lb}<\mathbf{x}<\mathbf{ub}\}$ 
		\State $\mathbf{x}'=\min \norm{\mathbf{A}\cdot\mathbf{x} - \mathbf{b}}^2$ for $\mathbf{x}\in\mathcal{F}$
		\State Set $\mathbf{x}=\mathbf{x}'$ for $\mathbf{x}\in\mathcal{F}$
		\EndWhile
	\end{algorithmic}
\end{algorithm}

The second procedure (\texttt{lsq\_lin}) solves general linearly-constrained quadratic programming problems like in \autoref{eq:sub_problem} using a modified version of the standard active-set technique described by \citet[algorithm~16.3]{Nocedal2006}. The approach consists of solving a sequence of equality-constrained linear least-squares problems, for which I follow \citet[algorithm~6.2.2]{Golub2013}, allowing for degenerate matrices using SVD. If the initial guess is unfeasible, I find a feasible point using the linear programming procedure \href{https://docs.scipy.org/doc/scipy/reference/generated/scipy.optimize.linprog.html}{\texttt{scipy.optimize.linprog}} and the \texttt{method='highs'} by \citet{Huangfu2017}. Alternatively, the quadratic sub-problem can be solved within \ppxf\ by minimizing the quadratic function in the form of \autoref{eq:quadratic} using a general quadratic programming solver. In the current implementation, I use the interior-point solver \texttt{solvers.coneqp} from the \textsc{cvxopt} package\footnote{Available from \url{https://cvxopt.org/}} by \citet{Andersen2011}, which is much faster for large-scale problems.
I have extensively tested the \texttt{lsq\_box} and \texttt{lsq\_lin} procedure described in this section by constructing batteries of tests using both exact analytic solutions and comparisons against \textsc{cvxopt}.

The \texttt{capfit} procedure has been the default nonlinear optimization algorithm in \ppxf\ for about three years and has been used as a general optimizer independently of \ppxf\ too. During this time \ppxf\ was used to fit millions of spectra from a variety of surveys (e.g. MaNGA \citealt{Bundy2015} and SAMI \citealt{Bryant2015}). This allowed me to fix its handling of rare failures in degenerate situations, which are difficult to encounter in idealized examples. When applied to unconstrained nonlinear least-squares problems \texttt{capfit} produces essentially the same iterates as the state-of-the-art LM implementations in \textsc{minpack} or \textsc{mpfit}, as expected. When used for box-constrained nonlinear least-squares problems \texttt{capfit} is generally at least as efficient as the best algorithms in \href{https://docs.scipy.org/doc/scipy/reference/generated/scipy.optimize.least_squares.html}{\texttt{scipy.optimize.least\_squares}}. However, \texttt{capfit} allows for the extra flexibility of using linear constraints, as well as for keeping variables tied to others or fixed.

\subsection{Setting linear constraints on the template weights}

In the previous section, I discussed the use of linear constraints on the kinematic parameters during the \ppxf\ fit. Here I note that linear constraints can be used also during the linear-fitting procedure in  \citet[sec.~3.3]{Cappellari2017}. These were used for example to constrain the sum of weights (e.g. the luminosity) of different template groups to constitute a certain fraction of the total light, e.g. to perform kinematic bulge/disks decompositions \citep[e.g.][]{Tabor2017,Tabor2019,Oh2020} or to study stellar population of different kinematic components \citep[e.g.][]{Shetty2020}.

\subsection{Multi-dimensional regularization}

\begin{figure}
	\includegraphics[width=\columnwidth]{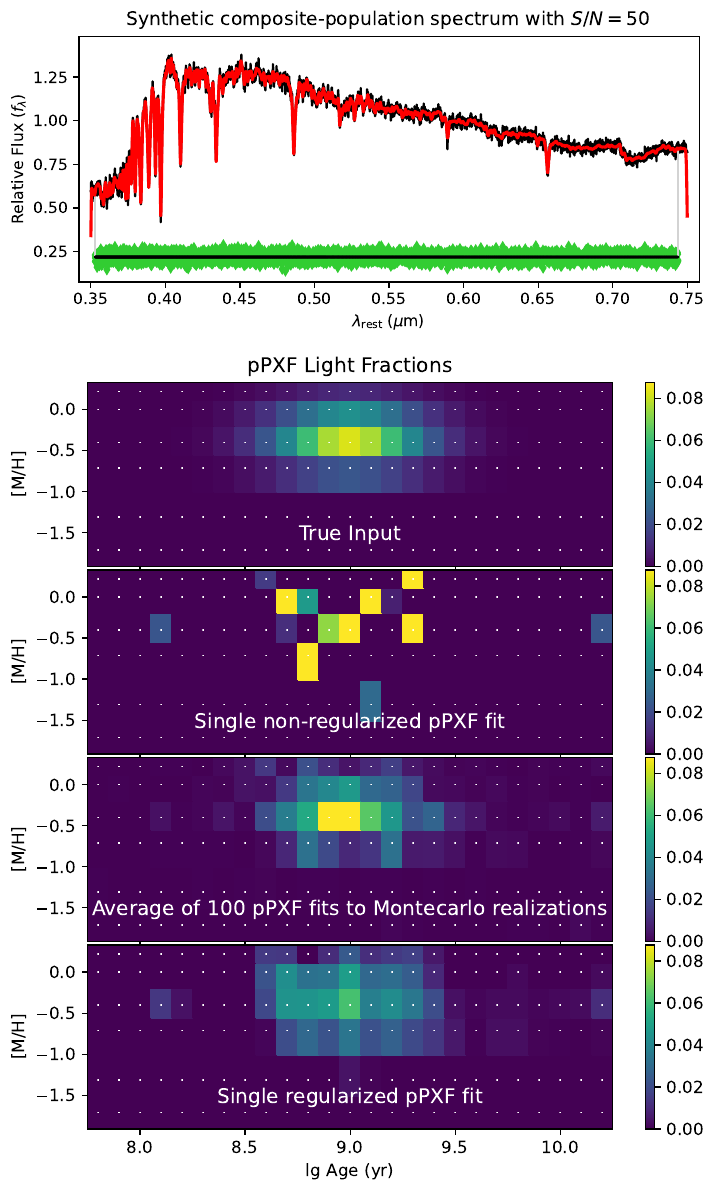}	
	\caption{Top panel: synthetic input spectrum with Gaussian noise at $S/N=50$. Second panel: input distribution of the $V$-band templates luminosity. Third panel: recovered weights from a single non-regularized \ppxf\ fit to one Monte Carlo realization of the noise. Fourth panel: average of the weights recovered with \ppxf\ by fitting 100 Monte Carlo realizations of the synthetic spectrum. Bottom panel: weights recovered with a single \ppxf\ fit with regularization \texttt{regul=30}. This approximates the average distribution as expected.
		\label{fig:montecarlo}}
\end{figure}

Considering without loss of generality that stellar population depends only on age, the fundamental equation used to model the spectrum of a composite stellar population is \citep[e.g.][sec.~2.3]{CidFernandes2005,Ocvirk2006,Conroy2013}
\begin{equation}\label{eq:sfh}
	G_{\rm mod}(\lambda)=\int_{t=0}^{t=T} {\rm SSP}_\lambda(t,Z)\cdot {\rm SFR}(T-t)\,  \dd t,
\end{equation}
where SFR is the star formation rate, ${\rm SSP}_\lambda$ is a Single Stellar Population spectrum per unit mass, with age $t$ and metallicity $Z$, while $T$ is the age of the Universe at the redshift of the galaxy. This expression is generalized in \ppxf\ to study the distribution of more parameters, like e.g.\ metallicity, $\alpha$ enhancement or IMF, in addition to the SFR. 

I pointed out in \citet[sec.~3.5]{Cappellari2017} that \autoref{eq:sfh} is an inhomogeneous Fredholm equation of the first kind, with kernel ${\rm SSP}_\lambda$. And the recovery of the ${\rm SFR}(t)$ from the observed $G_{\rm mod}$ is a textbook example of ill-conditioned inverse problem \citep[e.g.][sec.~19.0]{Hansen1998regul,Kabanikhin2011,Press2007}. This means that one cannot find a unique solution from real data without further assumptions.

In \citet[sec.~3.5]{Cappellari2017} I discussed the implementation of linear regularization \citep[e.g.][sec.~19.5]{Press2007} in \ppxf\ to address this issue and study the stellar population in galaxies. I gave a formula for the second-order one-dimensional regularization. Given that there are alternative ways of generalizing a measure of smoothness of a function in dimension larger than one \cite[e.g.][]{Brady1983}, I clarify here that the second-order regularization (\texttt{reg\_ord=2}) in \ppxf\ minimizes the total squared Laplacian $(\Delta w)^2$ of the weights $w$ distribution, while the first-order one (\texttt{reg\_ord=1}) minimizes the total squared gradient $(\nabla w)^2$. Both operators are implemented by standard finite differences.

\citet[19.4.1]{Press2007} point out that, under some sensible conditions, the regularized solution has a simple Bayesian interpretation: it represents the most likely solution for the weights, given an adjustable prior on the amplitude of the fluctuations. However, the meaning of the fundamental degeneracy of the stellar population inversion, as well as of regularization, is best illustrated with an example. 

I used a grid of 25 logarithmically-spaced ages $t$ and 6 metallicities $[M/H]$ from the SPS models by \citet{Vazdekis2015} to construct a synthetic spectrum in which the distribution of light contributed by each spectrum in the $V$ band follows a bivariate Gaussian distribution $\mathcal{N}(t,[M/H])$ with mean age $t_0=1$ Gyr, mean $[M/H]_0=-0.3$ and dispersion of 0.25 dex in both age and metallicity. I logarithmically sampled the spectrum at a velocity scale $\Delta V=c\,\Delta\ln\lambda=50$ \kms\ per spectral pixel.

I show the resulting spectrum in the top panel of \autoref{fig:montecarlo} and the input light-weights distribution in the second panel. The third panel shows a single \ppxf\ fit, which is characterized by discrete sharp peaks as expected due to the ill-conditioning of the inversion problem. The fourth panel shows the result of averaging the weights obtained by fitting with \ppxf\ 100 Monte Carlo realizations obtained by adding Gaussian noise on the same noiseless synthetic spectrum. Here, the average converges towards the true input distribution. Finally, in the bottom panel, I show the result of performing a single regularized \ppxf\ fit (with \texttt{reg\_order=2} and a typical \texttt{regul=30}). Here the distribution looks comparable to that of the average of multiple realizations.

Regularization has its limitations, in fact, it is by construction a trade-off between agreement with the data and smoothness \citep[e.g.][fig.~19.4.1]{Press2007}, which may introduce biases. In general, when one is one is obtaining results by averaging many spectra, it may be better not to use regularization, or only use a minimal amount, to reduce possible biases, while allowing the differences in the noise between spectra to act as Monte Carlo realizations. But regularization is very useful when interpreting individual spectral fits and even to reduce noise in the SPS models themselves, which may introduce spurious features in the solutions  (as I found later). 

One can use bootstrapping of the residuals, while repeating the \ppxf\ fits multiple times, to obtain averages as well as uncertainties in the distribution of the weights as done e.g. by \citet[figs.~8--13]{Kacharov2018}. In this case, it is important to perform the initial \ppxf\ fit, from which the residuals are extracted, using some regularization, to obtain a less noisy and more representative best-fitting spectrum. I achieved good results perturbing the residuals using the easy-to-use wild bootstrap method \citep{Davidson2008}.

\subsection{Global nonlinear fitting}

In the most common situations, e.g. when fitting a single stellar kinematic component with emission lines, the spectral fitting problem has a single global minimum and the local optimization method of \autoref{sec:capfit} is guaranteed to efficiently converge to it. However, in more complex situations, like when fitting multiple stellar or gas kinematic components, the fitting problem may present multiple minima and a local optimizer is not guaranteed to converge to the global minimum.

The standard way of dealing with multiple minima in \ppxf\ is to perform the optimization of the variables in which the $\chi^2$ function is multi-modal outside of \ppxf, while calling \ppxf\ with those variables fixed, from inside a wrapper function. For example, when studying multiple kinematic components one may sample a grid of velocities and call \ppxf\ with fixed velocities at every location \citep[e.g.][]{Mitzkus2017,Tabor2017,Bevacqua2022}. If one is interested in the full posterior of certain parameters, and computation time is not an issue, one may call \ppxf\ with those parameters fixed from within a Bayesian method like \textsc{MultiNest} \citep{Feroz2009}, \textsc{emcee} \citep{Foreman-Mackey2013emcee}, \textsc{AdaMet} \citep{Cappellari2013p15} or \textsc{dynesty} \citep{Speagle2020}, assuming the contribution of the non-fixed parameters to the posterior can be neglected.

\begin{figure}
	\includegraphics[width=\columnwidth]{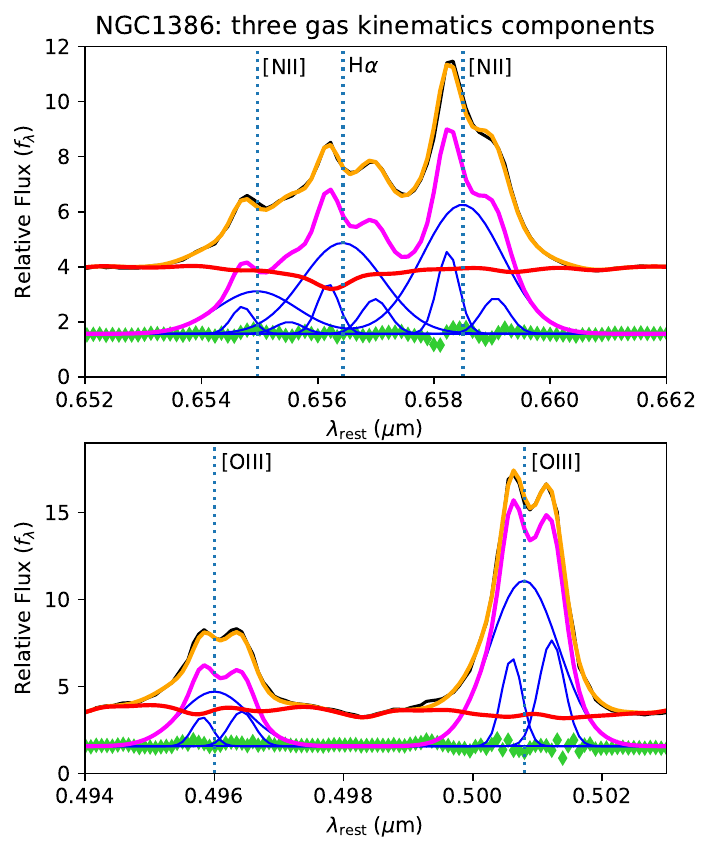}
	\caption{Fit of a MUSE spectrum of the active galaxy NGC~1386 (thin black line, mostly hidden by the fit), using the global-optimization option (\texttt{global\_search=True}) and linear constraints (\texttt{constr\_kinem}) in \ppxf. Each of the five emission lines is modelled with three kinematic components (see text). The orange line is the \ppxf\ total best fit, the red line is the best fitting stellar spectrum alone, while the magenta is for the gas emissions alone, with individual components shown in blue. The fit residuals, arbitrarily offset, are shown with green diamonds. I only plot the region with the key emissions, but I fitted the full optical spectrum, which is needed to constrain the underlying stellar contribution.
		\label{fig:muse_gas_fit}}
\end{figure}

In the current version of \ppxf\ one can also perform the global optimization within \ppxf. This is currently implemented using the function \href{https://docs.scipy.org/doc/scipy/reference/generated/scipy.optimize.differential_evolution.html}{\texttt{scipy.optimize.differential\_evolution}}, which uses the Differential Evolution algorithm by \citet{Storn1997}. The Scipy function allows for linear constraints using the method by \citet{Lampinen2002}. To save computation time, by default, I do not run the global optimization step until convergence, but I use it as starting point for the usual \textsc{CapFit} procedure.

An example of a situation where using both the global optimization and the linear constraints options can be useful, and the corresponding \ppxf\ fit is shown in \autoref{fig:muse_gas_fit}. The plots show the central spectrum of the active galaxy NGC~1386, extracted from the MUSE \citep{Bacon2010} integral-field spectroscopic observations presented in \citet{Venturi2021}. The emission line spectrum clearly requires at least three distinct kinematic components \citep[see also][]{Lena2015}. The definition of the three kinematics components may appear ill-defined, due to the extensive blending of the lines. However, a meaningful decomposition can be obtained with some simple assumptions. Here I required the kinematics $(V,\sigma)$ of all five emission lines to be the same within each of the three kinematic components. I additionally required the $\sigma_{\rm broad}$ of the broad component to be at least 200 \kms\ broader than either of the two narrow components as follows $\sigma_{\rm broad}>\sigma_{\rm narrow,1}$ and $\sigma_{\rm broad}>\sigma_{\rm narrow,2}$. These are linear constraints that I enforced using the \texttt{constr\_kinem} keyword in \ppxf. I also fix the ratios of the [OIII] and [NII] doublets to $1/3$.

Linear constraints in \ppxf\ were used extensively to produce the recent catalogue of broad and multiple gas emission line components for the full MaNGA galaxy survey \citep{Fu2023}.

\subsection{Fitting spectra and photometry}
	
Adding photometry to a full-spectrum fitting method is similar to adding a few extra pixels to the fit, which represent the fluxes measured in some observed photometric bands. The main differences are (i) that the photometric fluxes are independent of the line-of-sight velocity-distribution (LOSVD) $\mathcal{L}_n(v)$, unlike the spectroscopic ones and (ii) the photometry of a single galaxy is usually not enough to determine both the calibration errors and the template weights. This means that one cannot use polynomials as done for the spectroscopy.

I define a function that describes an individual template spectrum $T_n(\lambda)$ (either stars or gas), convolved $\ast$ with the LOSVD, which is allowed to be different for each of the $N$ templates
\begin{equation}
	g_n(\lambda)=T_n(\lambda) \ast \mathcal{L}_n(v).
\end{equation}
With this notation, the model for the galaxy spectrum becomes
\begin{align}\label{eq:model_spec}
G_{\rm mod}^{\rm spec}(\lambda) \,=\,&
\sum_{n=1}^{N} w_n \left[g_n(\lambda)\, A_n(\lambda)
\sum_{k=1}^{K} a_k\, \mathcal{P}_k(\lambda)\right]\nonumber \\
&+\, \sum_{l=0}^{L} b_l\, \mathcal{P}_l(\lambda)\, +\, \sum_{j=1}^{J} c_j\, S_j(\lambda),
\end{align}
where the $\mathcal{P}_k$ and $\mathcal{P}_l$ are multiplicative and additive polynomials respectively (of Legendre or trigonometric type) and the $S_j$ are optional spectra of the sky. This model is similar to the one in \citet[eq.~11]{Cappellari2017}, except that here each template spectrum can have a different attenuation function $A_n$. Moreover, both the attenuation and multiplicative polynomials can be used simultaneously, rather than being alternatives. This is especially useful when including photometry in the fit. 

The reason for this modification is that, when we have photometric bands that span a large wavelength range, we can infer the attenuation from the photometry itself, which is not modelled with polynomials. At the same time, we can use multiplicative polynomials to correct small errors in the spectral flux calibration. However, if we do not have photometry, we cannot tell apart reddening and multiplicative polynomials, because reddening is a special case of polynomials.

The model for the photometric measurements, in linear units which allow for negative fluxes, not magnitudes, is given by the following expression
\begin{equation}\label{eq:model_phot}
	G_{\rm mod}^{\rm phot}(\lambda_q) 
	= 	\sum_{n=1}^{N} w_n\, \langle g_n(\lambda)\, A_n(\lambda)\rangle_q,
\end{equation}
where $\langle \cdot \rangle_q$ represents the attenuated mean flux of the $n$-th template in the $q$-th photometric band with effective wavelength $\lambda_q$. Unlike the spectroscopic model of \autoref{eq:model_spec}, the photometric model of \autoref{eq:model_phot} does not include polynomials or the sky spectrum.

In the common case of a photon-counting or energy-integrating detectors, and assuming fluxes as $f_\lambda$ (e.g. in units of erg cm$^{-2}$ s$^{-1}$ \AA$^{-1}$) the mean flux is given by \citep[e.g.][eq.~A11]{Bessell2012}
\begin{equation}\label{eq:band_integ}
	\langle f_\lambda \rangle_q = 
	\frac{\int f_\lambda(\lambda) S_q(\lambda)\lambda\, d\lambda}
	{\int S_q(\lambda)\lambda\, d\lambda},
\end{equation}
where $S_q(\lambda)$ is the system photon response function and the integral extends over the region where $S_q$ is nonzero. The definition of mean flux in \autoref{eq:band_integ} is the one used in the standard definition of magnitudes in the ultraviolet (e.g for the GALEX spacecraft \citealt{Martin2005}), in the optical (e.g. for the SDSS optical survey \citealt{York2000}), or in the near-infrared (e.g. for the 2MASS survey \citealt{Skrutskie2006}). We can exactly convert mean fluxes in units of $f_\nu$  (for example, erg cm$^{-2}$ s$^{-1}$ Hz$^{-1}$) using this formula
\begin{equation}
	\langle f_\lambda \rangle_q=\langle f_\nu \rangle_q\frac{c}{\lambda_p^2},
\end{equation}
where $c$ is the speed of light and $\lambda_p$ the source-independent pivot wavelength defined as \citep[e.g.][eq.~A16]{Koornneef1986,Bessell2012} 
\begin{equation}
	\lambda_p^2=\frac{\int S(\lambda)\lambda\, d\lambda}{\int [S(\lambda)/\lambda]\, d\lambda}.
\end{equation}
One could use different definitions of the observed mean fluxes by simply replacing \autoref{eq:band_integ}.

In the common situation in which the covariance between the spectroscopic $G^{\rm spec}$ or photometric $G^{\rm phot}$ measurements are not known, or ignored, the residuals $\mathbf{r}$ from the fit are as in \citet[eq.~22]{Cappellari2017}
\begin{subequations}\label{eq:chi2_no_covar}
	\begin{align}
		&r_p = \frac{G_{\rm mod}^{\rm spec}(\lambda_p)-G^{\rm spec}(\lambda_p)}{\Delta G^{\rm spec}(\lambda_p)}, \quad p=1,\ldots,P\\
		&r_q = \frac{G_{\rm mod}^{\rm phot}(\lambda_q)-G^{\rm phot}(\lambda_q)}{\Delta G^{\rm phot}(\lambda_q)}, \quad q=P+1,\ldots,P+Q,
	\end{align}
\end{subequations}
with the difference that the vector of residuals now includes both the $P$ spectroscopic and the $Q$ photometric values. In other words, the total log-likelihood $\ln \mathcal{L}_{\rm total}$ of a fit now becomes the sum of the spectroscopic and photometric ones
\begin{equation}
\ln\mathcal{L}_{\rm total}=\ln\mathcal{L}_{\rm spec} + \ln \mathcal{L}_{\rm phot}
= -\frac{\chi^2_{\rm spec} + \chi^2_{\rm phot}}{2} + {\rm const.}
\end{equation}
Both the linear and nonlinear fit, the regularization and the possible treatment of covariances, proceed unchanged as already described in \citet[sec.~3.3--3.5]{Cappellari2017}. The only difference is one extra row in the matrix $\mathbf{A}$, defined in \citet[sec.~3.3]{Cappellari2017}, for every photometric measurement.

According to the mean value theorem for integration, for every $q$-th band and $n$-th template, there exists a wavelength  $\lambda_{q, n}$ which satisfies exactly 
\begin{equation}\label{eq:lam_eff}
	A_n(\lambda_{q,n})\, \langle g_n(\lambda) \rangle_q = \langle g_n(\lambda)\,A_n(\lambda) \rangle_q.
\end{equation}
When one has a good estimate of the galaxy redshift (e.g. from previous photometric redshift), or when performing a grid search for the best-fitting redshift with \ppxf, the redshift of the spectrum does not change much during each \ppxf\ fit. This makes the quantities $\langle g_n(\lambda) \rangle_q$ essentially independent of $\mathcal{L}_n(v)$. If I rewrite \autoref{eq:model_phot} as
\begin{equation}\label{eq:model_phot_approx}
	G_{\rm mod}^{\rm phot}(\lambda_q) 
	= \sum_{n=1}^{N} w_n\, A_n(\lambda_{q,n})\, \langle g_n(\lambda) \rangle_q,
\end{equation}
I can precompute the $\langle g_n(\lambda) \rangle_q$ and $\lambda_{q,n}$ for all templates before the fit, using the initial redshift estimate. With this approach, adding photometry to a fit takes almost no extra time compared to fitting only the spectrum.

\begin{figure}
	\includegraphics[width=\columnwidth]{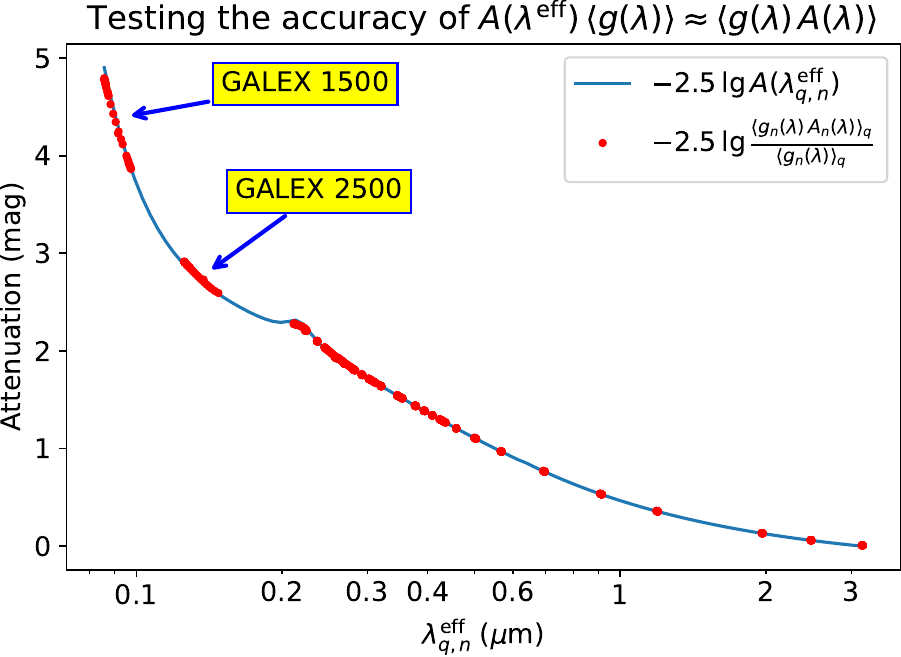}
	\caption{Comparison between the variation in the mean flux due to dust attenuation $\langle g_n(\lambda)\, A_n(\lambda)\rangle_q/\langle g_n(\lambda)\rangle_q$ (red circles) and the value of the attenuation curve at the rest-frame effective wavelength $A_n(\lambda_{q,n}^{\rm eff})$ (blue solid line), plotted versus the rest-frame effective wavelength $\lambda_{q,n}^{\rm eff}$ of each $q$-th filter and $n$-th template combination. This is computed for 28 filters at redshift $z=0.8$ and a set of templates spanning extreme ages and metallicities as described in the text. The $\lambda_{q,n}^{\rm eff}$ relative variation and the corresponding variation of the attenuation at that wavelength for a given filter are only significant in the far ultraviolet GALEX filters (indicated by the blue arrows). Even in that case, $A_n(\lambda_{q,n}^{\rm eff})$ accurately predicts the true integrated attenuation. \label{fig:attenuation_approximation}}
\end{figure}
	
I found that the flux-weighted effective wavelength \citep[e.g.][eq.~A21]{Bessell2012}
\begin{equation}\label{eq:lam_eff_approx}
	\lambda_{q,n}^{\rm eff} = 
	\frac{\int g_n(\lambda) S_q(\lambda)\lambda^2\, d\lambda}
	{\int g_n(\lambda) S_q(\lambda)\lambda\, d\lambda},
\end{equation}
well approximates the wavelength $\lambda_{q,n}$ defined by \autoref{eq:lam_eff}, for a range of attenuation parameters.	
In \autoref{fig:attenuation_approximation} I illustrate how accurately \autoref{eq:lam_eff} is verified when approximating $\lambda_{q,n}\approx\lambda_{q,n}^{\rm eff}$. I used all the 28 photometric bands described in \autoref{sec:photometry} and all 387 \textsc{fsps} SPS templates introduced in \autoref{sec:ppxf_setup}, which span extreme ranges of age and metallicity. I adopt the attenuation function of \autoref{eq:attenuation}, with realistic parameters $(A_V, \delta, E_b, f_{\rm nodust})=(1, 0, 0.25, 0)$ and the median redshift $z=0.8$ of the LEGA-C sample. For every template-band combination, I compare the rigorous variation in mean flux due to the attenuation of each $n$-th template in the $q$-th band $\langle g_n(\lambda)\, A_n(\lambda)\rangle_q/\langle g_n(\lambda)\rangle_q$ with the value of the attenuation curve at the effective wavelength $A_n(\lambda_{q,n}^{\rm eff})$. The two quantities must agree when the photometric band is narrow or the attenuation is approximately constant within the band, as is generally the case at optical or near-infrared wavelengths. However, in the far ultraviolet, in the GALEX bands, $\lambda_{q,n}^{\rm eff}$ and the corresponding $A_n(\lambda_{q,n}^{\rm eff})$ vary significantly for different templates in the same band. Even so, as shown in \autoref{fig:attenuation_approximation} the approximation is still much better than our uncertainty of the attenuation curve itself. Moreover, for these large attenuations, generally little flux is detected in the far ultraviolet, making the observed uncertainties very large. For these reasons, in the analysis presented here, I use \autoref{eq:model_phot_approx} to model the attenuation on the photometry. When higher accuracy is required the full expression of \autoref{eq:band_integ} can be used.

\subsection{Dust attenuation}

As shown in \autoref{eq:model_spec} and \autoref{eq:model_phot}, the new \ppxf\ method allows each template to have a different attenuation curve. This feature can be used to vary the attenuation curve for specific groups of templates, based on the current understanding of dust attenuation in galaxies \citep[see review by][]{Salim2020}. Three groups of attenuation curves are expected to be useful: (i) for very young stars (with ages $t\la10^7$ yr), which are still embedded in their birth clouds  \citep{Charlot2000,Granato2000}; (ii) for the entire stellar population (both young and old), due to diffuse dust; and (iii) for the gas emission lines from star-forming regions.

In \ppxf\ one can adopt a generic function, which can be different for different templates and can have an arbitrary number of parameters. The parameters can have bounds or can be kept fixed. By default I currently implemented a four-parameters attenuation function in linear units $A(\lambda)=f(A_V, \delta, E_b, f_{\rm nodust})$ defined by
\begin{subequations}\label{eq:attenuation}
	\begin{align}
		&D(\lambda)=\frac{E_b\,(\lambda\,\Delta\lambda)^2}{(\lambda^2 - \lambda_0^2)^2 + (\lambda\,\Delta\lambda)^2}\label{eq:dust1}\\
		&k(\lambda) = \frac{A_V}{R_V}\left[k'(\lambda)+D(\lambda)\right]\left(\frac{\lambda}{\lambda_V}\right)^\delta\label{eq:dust2}\\
		&A(\lambda) = f_{\rm nodust} + (1 - f_{\rm nodust})\,10^{-0.4\, k(\lambda)}\label{eq:dust3}.
	\end{align}
\end{subequations}
Here \autoref{eq:dust1} is the Lorentzian-like Drude function adopted by \citet{Noll2009} to describe the UV bump around $\lambda_0=0.2175$ \micron, with width $\Delta\lambda=0.035$ \micron. The \autoref{eq:dust2} is the expression adopted by \citet{Kriek2013}, which includes the attenuation $k'(\lambda)$ and $R_V=4.05$ from \citet[eq.~4 and 5]{Calzetti2000}, and allows for a variable UV slope $\delta$ around the pivot $V$-band wavelength $\lambda_V=0.55$ \micron. Optionally, one can make $E_b$ a function of $\delta$ \citep[eq.~3]{Kriek2013}
\begin{equation}
	E_b = 0.85 - 1.9\times\delta.
\end{equation}
Finally \autoref{eq:dust3} allows one to specify the fraction $f_{\rm nodust}$ of the stellar population (for the given template) that is unattenuated, as suggested by \citet{Lower2022}. The resulting $A(\lambda)$ is the factor to multiply the template at the given wavelength to model the attenuation effect. 

\section{Data and SPS models}
\label{sec:data}

In the rest of this paper, I present an analysis of the combined photometric and spectroscopic data, for a sample of about 3200 galaxies at $0.6<z<1$, making use of various of the new features of \ppxf\ introduced in the first part of the paper. 

\subsection{Spectroscopy}
\label{sec:spectroscopy}

I study a subset of the galaxy sample of the LEGA-C survey \citep{vanderWel2016}. It is an ESO ESO/Very Large Telescope (VLT) public spectroscopic survey targeting galaxies in the redshift range $0.6 < z < 1$, selected based on their observed $K_s$-band luminosity in the UltraVISTA/COSMOS catalogue by \citet{Muzzin2013}, with a small variation of the $K_s$ limit with $z$. In this work, I use the data from the LEGA-C third data release (DR3) presented in \citet{vanderWel2021}. This redshift selection results in a mass-complete sample of 3445 galaxies in DR3 (90\% completeness above $\lg(M_\ast/\msun)\ga10.3$). The selection, the completeness level, the characteristics of the sample and the data reduction are discussed extensively in \citet{vanderWel2016} and \citet{Straatman2018}. 

For my study, I focused only on the subsample of 3197 galaxies (including duplicates) in the DR3 catalogue with spectroscopic redshift $0.6 < z < 1$ and with measured stellar velocity dispersion $\sigma_\ast$. This a sample has redshift $z=[0.67, 0.76, 0.93]$ and average $S/N=[7,14,26]$ per \AA\ at the 16th ($-1\sigma$), 50th (median) and 84th ($+1\sigma$) percentiles. However, I verified that all my results are unchanged if I restrict the sample to the redshift range to $0.7 < z < 0.9$.

The survey data consist of spectra observed with the VLT/VIMOS multi-object spectrograph \citep{LeFevre2003}  covering the wavelength range 0.63-0.88 \micron\ with a spectral resolution $R\approx3500$, equivalent to an instrumental dispersion $\sigma_{\rm inst}\approx36$ \kms\ \citep{vanderWel2021}. For the \ppxf\ fits I logarithmically rebinned the spectra to a velocity scale $\Delta V=c\Delta\ln\lambda=\sigma_{\rm inst}$ \citep[eq.~8]{Cappellari2017} to make sure the spectrum is Nyquist sampled.

\subsection{Photometry}
\label{sec:photometry}

I use two large collections of photometric measurements for the LEGA-C galaxies. The first is the UltraVISTA/COSMOS catalogue by \citet{Muzzin2013}. It includes PSF-matched photometry in 30 bands from 0.15 to 24 \micron. The catalogue is based on the $YJHK_s$ NIR imaging data from UltraVISTA \citep{McCracken2012}. The optical data consist of broad-band Subaru/SuprimeCam data ($g^+r^+i^+z^+B_jV_j$), as well as $u^\ast$ data from the CFHT/MegaCam \citep{Taniguchi2007,Capak2007a}. It also includes the 12 optical medium bands (IA427–IA827) from Subaru/SuprimeCam \citep{Capak2007a}. Also included are the GALEX FUV and NUV channels \citep{Martin2005}, and the 3.6 \micron\, 4.5 \micron\, 5.8 \micron\, 8.0 \micron\, and 24 \micron\ channels from Spitzer’s IRAC+MIPS cameras \citep{Sanders2007}. The model predictions within the FUV GALEX band at $z\sim0.8$ are quite uncertain and few photons are generally expected to escape at those wavelengths. However, I still include this band in the fit to verify that this is indeed the case in the data. Moreover, significant detections are observed for the most star-forming galaxies.

The second photometric catalogue is the COSMOS2020 by \citet{Weaver2022}. Highlights of this catalogue, compared to the one by \citet{Muzzin2013}, are much deeper Subaru Hyper Suprime-Cam $grizy$ broadband photometric measurements \citep{Aihara2019} and deeper UltraVISTA DR4 observations $YJHK_s$. The extra depth is not an important feature for my study, as the LEGA-C galaxies were all well-detected in \citet{Muzzin2013} by design. However, I use the COSMOS2020 to assess the sensitivity of my results to the use of independent datasets. For this work, I adopt the catalogue produced with the \textsc{farmer} profile-fitting photometric extraction tool. For both photometric catalogues, I only included the typically 28 bands which have a transmission FWHM fully contained in the wavelength range of the adopted stellar population templates (see later) at the redshift of each galaxy.

The current COSMOS2020 \textsc{farmer} catalogue does not have the two GALEX bands, so I added them to compare it more closely with the UltraVISTA/COSMOS catalogue. I used the following steps for each galaxy: (i) I selected the bands that were common between the COSMOS2020 and the UltraVISTA/COSMOS catalogues. (ii) I used \autoref{eq:scaling} to perform a linear least-squares fit and find a normalization factor $\kappa$ that matches the two photometries for that galaxy. (iii) I applied the same factor $\kappa$ to scale the two GALEX bands and included them in the COSMOS2020 bands.

\subsection{Stellar population synthesis models}
\label{sec:ppxf_setup}

I used three independent SPS models to assess the sensitivity of the results to some of the adopted model assumptions. I selected the models based on two criteria (i) the ability to generate model spectra from the far UV at 0.1 \micron\ to about 2 \micron, to be able to constrain the decrease of $f_\lambda$ towards the NIR region of the galaxy spectra and (ii) to include model spectra down to a young age of 1 Myr, to reproduce the many actively star-forming galaxies that are present in the sample. The age criterion forces me to exclude from this study the models by Vazdekis and Maraston, which I have extensively used in the past.

The three SPS models that satisfy my requirement and that I adopted are (i) the \textsc{fsps}\footnote{Available from \url{https://github.com/cconroy20/fsps}} \citep{Conroy2009,Conroy2010}, (ii) the \textsc{galaxev}\footnote{Available from \url{http://www.bruzual.org/bc03/}} \citep{Bruzual2003} and (iii) the \textsc{Bpass}\footnote{Available from \url{https://bpass.auckland.ac.nz/}} SPS models \citep{Stanway2018,Byrne2022}. 

For all three models, I tried to select a consistent set of templates. In all cases, I adopted the same set of 43 ages logarithmically spaced by 0.1 dex from 1 Myr to 15.85 Gyr, defined as 
\begin{equation}
	\lg ({\rm Age}/{\rm yr}) = 6, 6.1, 6.2,\ldots,10.2.
\end{equation}
The oldest age is 2--3$\times$ older than the age of the Universe at the redshift of the sample, which varies in my standard cosmology from 5.75 -- 7.75 Gyr between $z=1$ -- 0.6, but I did not truncate the models to physical ages, to check how well the data themselves can constrain the galaxy ages. For comparison, I additionally ran models where I constrained the maximum age in the fit to each galaxy to the age of the Universe at its redshift.
I also excluded from all models the most extreme low metallicities $[Z/H] < -2$. Here is my other setup for the three SPS:
\begin{enumerate}
	\item \textbf{The \textsc{fsps} models} allow one to compute SPS models for a specified set of parameters. I used the Python bindings\footnote{Available from \url{https://github.com/dfm/python-fsps}} \citep{Johnson2021a} and the latest public v3.2 to compute a set of spectra with the above ages and 9 equally-spaced metallicities $[Z/H]=[-1.75, -1.5, -1.25, -1., -0.75, -0.5, -0.25, 0, 0.25]$, for a total of 387 SPS templates. I adopted a \citet{Salpeter1955} IMF with a lower/upper mass cut of 0.08 and $100$ \msun\ respectively, for consistency with \textsc{bpass}, and used the MIST isochrones \citep{Choi2016}. But I note that my results are virtually insensitive to the slope of the IMF at lower masses. I computed the SPS without including the effect of gas or dust and adopted default parameters for the other parameters. This returns SPS spectra computed using the MILES stellar library  \citep{Sanchez-Blazquez2006,FalconBarroso2011miles} for the optical region, which is the one I fit in the LEGA-C spectra.
	
	\item \textbf{The \textsc{galaxev} models} provide a Fortran code (version 2020) which I used to produce a set of SPS spectra with the same ages as above and computed at the provided 5 metallicities $[M/H]=[-1.74, -0.73, -0.42, 0, 0.47]$, for a total of 215 SPS templates. This SPS model also uses the MILES library to generate spectra of the optical region. Also here I adopted a Salpeter IMF. The models use the Padova isochrones \citep{Bertelli1994,Girardi2000,Marigo2008}.
	
	\item \textbf{The \textsc{bpass} models} v2.3 are provided as a set of files precomputed at a given set of metallicity. I adopted the 10 metallicities\footnote{The models are specified in metal mass fraction $Z$, and I converted it to $[Z/H]$ with $\sim10$ \% accuracy.} $[Z/H]=[-1.3, -1, -0.8, -0.7, -0.5, -0.4, -0.3, 0, 0.2, 0.3]$, for a total of 430 SPS templates. The models are provided for a single IMF having a power slope -2.35 (Salpeter slope) above $M>0.5$ \msun\ and a slope -1.3 at lower masses. I used the version of the SPS for single stars, ignoring binaries, with $[\alpha/Fe]=0$, for consistency with the other two SPS models. These SPS models are fully synthetic. They use isochrones produced by a derivative of the Cambridge \textsc{stars} code \citep{Eggleton1971} as described by \citet{Eldridge2008}.
\end{enumerate}

\section{Dynamical Masses from Sersic Photometry}
\label{sec:jam_of_sersic}

We don't know the true masses of galaxies, so we can't tell how good the mass estimates from different stellar population codes are. One way to test the accuracy is to use good dynamical models. I show how I do this in this section. I use mass-follow-light axisymmetric JAM dynamical models which are as quick and simple to use as the usual virial estimator \citep[e.g.][]{Cappellari2006}, and require the same data, but do not have its problems.

\subsection{Updated coefficient for the Sersic profile}

\begin{figure}
	\includegraphics[width=\columnwidth]{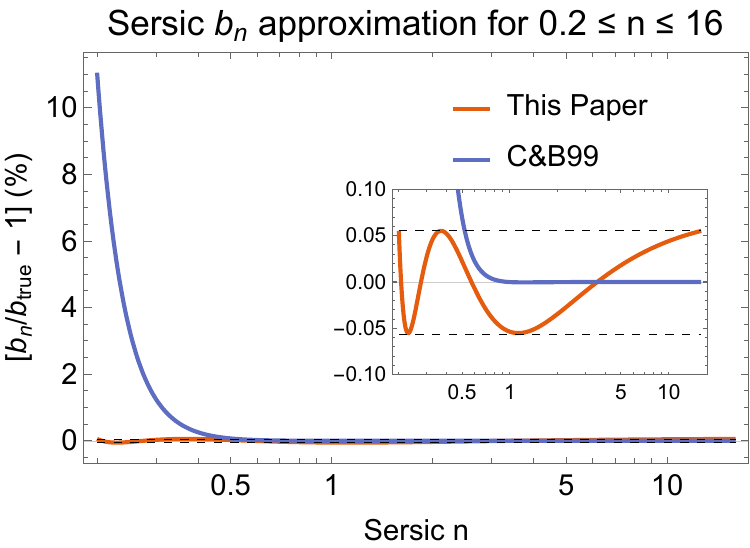}
	\caption{Fractional difference between the true coefficient $b(n_{\rm Ser})$ in \autoref{eq:sersic} and the minimax approximation (red line) for the interval $0.2<n_{\rm Ser}<16$ of \autoref{eq:bn_approx}. I also show for comparison the approximation (blue line) by \citet{Ciotti1999}, which is very accurate for $n_{\rm Ser}>0.5$, but starts deviating rapidly at smaller $n_{\rm Ser}$. The inset shows the same thing using a different scale for the $y$-axis.  \label{fig:sersic_coefficient}}
\end{figure}

I assume a galaxy with surface brightness described by a Sersic profile with elliptical isophotes of constant axial ratio $q^{\rm Ser}_{\rm obs}$
\begin{equation}\label{eq:sersic}
	I(m)=I(0)\, \exp\left[-b(n_{\rm Ser})\,
	\left(\frac{m}{R_{\rm e}^{\rm Ser}}\right)^{1/n_{\rm Ser}}\right],
\end{equation}
where the elliptical radius is
\begin{equation}
	m^2 = x^2 + \left(\frac{y}{q^{\rm Ser}_{\rm obs}}\right)^2,
\end{equation}
with $x$ aligned along the galaxy photometric projected major axis.

The parameter $b(n_{\rm Ser})$ is defined by the requirement that $R_{\rm e}^{\rm Ser}$ represents the semi-major axis of the isophote containing half of the total luminosity of the Sersic model. This implies that $b(n_{\rm Ser})$ is the solution of $\Gamma(2\,n_{\rm Ser})=2\,\Gamma(2\, n_{\rm Ser}, b(n_{\rm Ser}))$ \citep{Ciotti1991}, where $\Gamma(a,z)$ is the incomplete gamma function \citep[\href{https://dlmf.nist.gov/8.2.E2}{equation~8.2.2}]{Olver2010nist} and $\Gamma(a)=\Gamma(a,0)$ is the complete one \citep[\href{https://dlmf.nist.gov/5.2.E1}{equation~5.2.1}]{Olver2010nist}. A useful approximation for $b(n_{\rm Ser})$ was presented by \citet{Ciotti1999}. This is very accurate for values of $n_{\rm Ser}\ga0.5$ but starts becoming rapidly inaccurate for smaller $n_{\rm Ser}$ values. The LEGA-C catalogue contains values of the Sersic index down to $n_{\rm Ser}=0.2$ where I found that the \citet{Ciotti1999} approximation for $b(n_{\rm Ser})$ reaches an error of 11\% (\autoref{fig:sersic_coefficient}). 

To overcome this limitation I computed an alternative approximation for $b(n_{\rm Ser})$. I adopted the same number of terms and the same mathematical form, but I adjusted the coefficients to obtain the minimax solution, minimizing the maximum absolute relative error of $b(n_{\rm Ser})$ using nonlinear optimization over an interval including the most extreme Sersic indices existing in the literature. I found the following expression
\begin{equation}\label{eq:bn_approx}
b(n_{\rm Ser}) = \frac{0.000207}{n_{\rm Ser}^2} + \frac{0.015987}{n_{\rm Ser}} - 0.34025 + 2.0015\times n_{\rm Ser},
\end{equation}
which has a maximum absolute relative error of $5.6\times10^{-4}$ in the whole interval $0.2\le n_{\rm Ser}\le 16$. The fact that the relative error reaches its maximum value, with alternating sign, at five values of $n_{\rm Ser}$  (\autoref{fig:sersic_coefficient}) confirms that this is indeed the minimax solution for the adopted function and interval \citep[e.g.][sec.~5.15]{Press2007}.

\subsection{Jeans Anisotropic Models from Sersic photometry}
\label{sec:jam}

I performed dynamical modelling of the full LEGA-C sample of 3197 galaxies with measured $\sigma_*$ and $0.6<z<1$ using the Jeans Anisotropic Modelling (JAM) method\footnote{I used v7.2 of the \textsc{JamPy} Python package  \url{https://pypi.org/project/jampy/}} \citep{Cappellari2008,Cappellari2020}.

Dynamical modelling of LEGA-C galaxies with JAM was previously done for a subset of 797 galaxies by \citet{vanHoudt2021} using a Bayesian approach, but I extended it to all galaxies with measured $\sigma_*$. A simpler alternative to dynamical models would be to use a virial estimation of the galaxy masses \citep[e.g.][]{Cappellari2006} based on the fitted parameters of the \citep{Sersic1968} profiles provided in the LEGA-C DR3 catalogue \citep{vanderWel2021}. The virial estimates are also included in the DR3 catalogue.

The main advantage of virial estimators is that they are fast and easy. However, a common limitation is that they do not account for differences in the spectroscopic aperture and the instrumental point-spread function, which can be significant, especially in high-redshift observations like LEGA-C. Moreover, virial estimators do not allow one to explicitly assume an intrinsic shape or anisotropy for the galaxies under study.

Although virial estimators are still useful to study scaling relations \citep[e.g.][]{Cappellari2013p15,Li2018,Zhu2023a}, nowadays there is no reason to use them to compute galaxy masses. In fact, using e.g. the public JAM package, one can compute a more reliable dynamical mass for a galaxy approximated by a Sersic model with a similar time and effort as using the virial estimator, while allowing for intrinsic shape, anisotropy, aperture and PSF effects, without introducing unnecessary approximations.

To simplify this task of computing accurate dynamical masses of galaxies with available fitted Sersic parameters and $\sigma_*$, I developed a simple procedure \textsc{jam\_axi\_sersic\_mass} and I have made it publicly available in the updated version 7.2 of the JAM package. The procedure requires the following inputs from the users:
\begin{enumerate}
\item The parameters of the Sersic model for a galaxy: 
	\begin{enumerate}
		\item The semi-major axis of the half-light isophote $R_{\rm e}^{\rm Ser}$,
		\item The Sersic index $n_{\rm Ser}$, and 
		\item The observed axial ratio of the isophotes $q^{\rm Ser}_{\rm obs}$.
	\end{enumerate}
\item The assumptions about the intrinsic properties of the galaxy:
	\begin{enumerate}
		\item The intrinsic axial ratio $q^{\rm Ser}_{\rm intr}$ and 
		\item The typical orbital anisotropy $\beta$.
	\end{enumerate}
\item The parameters of the spectroscopic observations: 
	\begin{enumerate}
		\item The size and shape of the spectroscopic aperture, and 
		\item The parameters of the PSF.
	\end{enumerate}
\item The observed second moment $\sigma_*$ (which includes both rotation and random motions) and uncertainty of the stellar line-of-sight velocity distribution, preferably measured at a similar wavelength as the imaging used to fit the Sersic model.
\end{enumerate}
Given an angular diameter distance $D_A$, the procedure then returns the dynamical mass $M_{\rm JAM}$ of the Sersic model and its formal uncertainty in a fraction of a second.

The procedure uses the method and \textsc{mge\_fit\_1d} routine within the \textsc{MgeFit} package\footnote{V5.0 of Python \textsc{MgeFit} from \url{https://pypi.org/project/mgefit}} \citep{Cappellari2002mge} to accurately fit the one-dimensional Sersic profile of \autoref{eq:sersic} with a one-dimensional Multi-Gaussian Expansion (MGE). Then assumes a fixed axial ratio $q^{\rm Ser}_{\rm obs}$ for all MGE Gaussians and an arbitrary reference total mass $M_0$ for the model. It uses the \textsc{jam\_axi\_proj} procedure in the \textsc{JAmPy} package \citep{Cappellari2008,Cappellari2020} to calculate a PSF-convolved prediction for the $V_{{\rm rms},j}$ at a large set of discrete sky locations $(x_j,y_j)$ finely sampling the adopted spectroscopic aperture. The luminosity-weighted second moment inside the whole aperture is computed as
\begin{equation}
	V_{\rm rms}^2=\frac{\sum_j I_j V^2_{{\rm rms},j}}{\sum_j I_j},
\end{equation}
where the summation extends to the pixels of flux $I_j$ inside the aperture.
Given the general scaling $M\propto V^2$ between the total mass and velocities in a dynamical model, the dynamical mass of the Sersic model is then given by
\begin{equation}
	M_{\rm JAM}=\frac{M_0\, \sigma_*^2}{V_{\rm rms}^2}.
\end{equation}

The physical meaning of the dynamical mass $M_{\rm JAM}$, as derived from mass-follow-light dynamical models or virial estimators, is often a source of confusion. This is because $M_{\rm JAM}$ is neither a total stellar mass, nor a total mass of a galaxy which includes its dark halo. Moreover, the value of $M_{\rm JAM}$ is highly dependent on the extrapolated outer profile. For example, a Sersic model with $n_{\rm Ser}=6$ contains 21\% of its total light outside $4R_{\rm e}$, which is about the maximum radius one can observe in typical photometry. This strong dependence on extrapolation prevents comparison of masses when accuracies better than a few 20\% are desired.

The quantity that both dynamical and stellar population models are robustly measuring is the mass-to-light radius within the region covered by the spectroscopic (for the dynamics) or photometric (for the population) observations. Specifically, if one divides the Sersic dynamical mass $M_{\rm JAM}$ returned by the procedure by the analytic total luminosity of the same Sersic model \citep{Ciotti1991}
\begin{equation}
	L_{\rm Ser} = \pi\,  I(0)\, \Gamma(2 n_{\rm Ser}+1)\,  \left(\frac{R_{\rm e}^{\rm Ser}}{b^{n_{\rm Ser}}}\right)^2 q^{\rm Ser}_{\rm obs},
\end{equation}
the mass-to-light ratio
\begin{equation}
	(M/L)_{\rm JAM}=M_{\rm JAM}/L_{\rm Ser}
\end{equation}
provides a very accurate approximation of the average $M/L$ within a sphere of radius comparable to the size of the spectroscopic aperture.

I used \textsc{jam\_axi\_sersic\_mass} to compute $M_{\rm JAM}$ and $(M/L)_{\rm JAM}$ for all the LEGA-C galaxies in my subsample. I assumed a spectroscopic aperture of $1''\times1''$ and a characteristic PSF of 0\farcs75 FWHM from \citet{vanHoudt2021}. I adopted as the intrinsic axial ratio for all galaxies the mean value $q^{\rm Ser}_{\rm intr}=0.41$ of the Gaussian distribution inferred by \citet{vanHoudt2021} by inverting the observed shape distribution of the LEGA-C sample. For the anisotropy, I used a typical value $\beta=0.2$ expected for the assumed mean intrinsic shape \citep[fig.~9]{Cappellari2016}. I assumed a cylindrically-oriented (\texttt{align='cyl'}) velocity ellipsoid for JAM \citep{Cappellari2008}. The parameters of the Sersic profiles $R_{\rm e}^{\rm Ser}$, $n_{\rm Ser}$, $q_{\rm obs}^{\rm Ser}$ and $L_{\rm Ser}$ come from the LEGA-C DR3 catalogue \citep{vanderWel2021}.

I also used the dynamical models to calculate the average projected density $\Sigma_1^{\rm JAM}$ within a circle of radius $R=1$ kpc at the angular diameter distance of the galaxy. This quantity was shown to closely relate to galaxy quenching \citep{Cheung2012,Fang2013}, like $\sigma_*$. However, here instead of using the stellar mass as in previous studies, I used the dynamical mass from JAM. I obtained this value by circularising and analytically integrating the best fitting MGE as described in equation~(11) of \citet{Cappellari2013p15}.

\section{Setup for \ppxf\ and tests}

In this study, fitting (twice) the VIMOS spectrum ($\approx2800$ spectral pixels) and typically 28 photometric bands for a single galaxy with \ppxf\ takes about 1 min. This compares with the ``roughly 100 CPU hours'' reported by \citet{Tacchella2022} in a similar state-of-the-art study using DEIMOS spectroscopy and the \textsc{prospector} Bayesian code \citep{Johnson2021}. This is a computation time difference of nearly four orders of magnitude! Of course, the two methods perform quite different tasks and the large computational cost is a standard feature of Bayesian methods and not a weakness by itself. However, the execution time of a method affects the kind of tasks one can address and the variety of modelling choices one can explore as I outline in this section.

\subsection{Non-parametric population model}

A key feature made possible by least-squares methods is the ability to explore non-parametrically the joint distribution of SFH and chemical composition with high resolution. My setup uses 43 non-parametric age bins, and each age bin is allowed to have a different non-parametric metallicity (5 -- 10 bins depending on the SPS code) for a total of up to 430 bins. This contrasts with the 10 non-parametric age bins and the single metallicity for the entire galaxy adopted by \citet{Tacchella2022}. Crucially, even when using a non-parametric grid of a few hundred templates the least-squares method \emph{guarantees} global convergence to the most likely weights distribution. This is because the constrained quadratic-programming problem being solved \citep[eq.~27]{Cappellari2017} is known to possess a unique global minimum \citep[e.g.][]{Nocedal2006}. This contrast with Bayesian methods, where \citet[sec.~3.2]{Tacchella2022} reported that ``the fits do not converge within a reasonable amount of time'' with 14 non-parametric bins. The use of non-parametric models is important for a proper recovery of the stellar population in galaxies \citep{Lower2020}.

\subsection{Polynomials}

A least-squares method like \ppxf\ allows for a quick exploration of different modelling assumptions. In the course of this study, I was able to easily test different options, each with all three SPS models, for all 3200 galaxies, to test how they affected the final results. To fit the spectrum, we can choose between additive or multiplicative polynomials (the photometry does not use any polynomials). Additive polynomials are specified by the \ppxf\ keyword \texttt{degree} and can help with template mismatch, AGN modeling or sky subtraction errors. Multiplicative polynomials are specified by the \ppxf\ keyword \texttt{mdegree} and can account for spectral flux calibration issues or reddening effects.

For my tests, I run models with both multiplicative and additive polynomials degree from \texttt{mdegree=degree=-1} (i.e. using only attenuation and no polynomials) to degree 4 and found that the solution changes slightly without polynomials but quickly stabilizes as soon as one allows for a nonzero degree. Results were similar when only using additive or only multiplicative polynomials to adjust the spectral stellar continuum. This was a non-obvious result in the present analysis, given that the LEGA-C spectra were calibrated using SPS models \citep{vanderWel2021} and this non-standard calibration may leave an influence on the results. The polynomials should effectively remove any memory of possible inaccuracies in spectral calibration. I adopted the \ppxf\ keywords \texttt{mdegree=2, degree=-1} for my standard setup.

\subsection{Dust attenuation model}
\label{sec:dust}

I truncated all three the SPS templates to $0.01\,\micron<\lambda<5\,\micron$ (except for the \texttt{bpass} SPS which only extend to $\lambda<2\,\micron$) to remove the influence of dust on the spectral shape \citep[e.g.][fig.~1]{Conroy2013}. This is because the modelling of dust from the energy balance of UV light reradiated to the IR requires several further assumptions and is not implemented in all modelling codes. Moreover, the currently available bands are included in the fitted range anyway. This situation is changing rapidly with the James Webb Space Telescope (JWST) and will soon be revisited.

For the full set of 3200 galaxies, I experimented with three different assumptions for the attenuation: (i) I adopted a single four-parameter attenuation for all stellar templates as in \autoref{eq:attenuation}; (ii) I reduced the attenuation curve to two parameters $(A_V,\delta)$ by assuming $f_{\rm nodust}=0$ and adopting the $E_b-\delta$ relation by \citet[eq.~3]{Kriek2013}; I still applied this attenuation for all stellar templates; (iii) I adopted the two-components attenuation model by \citet{Charlot2000}. For this, I used a birth-cloud attenuation of the form 
\begin{equation}
	k(\lambda) = A'_V\left(\frac{\lambda}{0.55\,\micron}\right)^{-1}.
\end{equation}
I apply this only to the stellar templates younger than 10 Myr, while I used the same attenuation as in (ii) for the diffuse dust component affecting all stellar templates. In all three cases, I fit a different \citet{Calzetti2000} attenuation curve (\autoref{eq:attenuation} with $E_b=\delta=f_{\rm nogas}=0$) for the gas emission lines templates.

I found that options (i) and (ii) produce an insignificant difference in the final results, but in the first case there is more degeneracy in the attenuation parameters, which makes any trend between dust parameters less obvious. Option (iii) generates a similar result for the older and intermediate populations, as expected. However, when allowing the youngest population to have its own attenuation, this becomes completely degenerate with the amount of star formation in that same young component. As a result, one can obtain good fits with an unlikely large attenuation associated with an equally large star formation, but with complete degeneracy between the two parameters. This is the well-known SFR-dust degeneracy mentioned in \autoref{sec:intro}. It can be broken by introducing extra assumptions on the dust geometry and reradiated UV fraction, combined with rest-frame IR data, which I do not have. 

In conclusion, I adopted as my standard choice the two-parameters attenuation function of option (ii), which applies to the whole population, as done e.g. by \citet{Kriek2013}. I enforced bounds on the parameters as $-1<\delta<0.4$ and $0<A_V<4$. As illustrated by \citet[fig.~1]{Kriek2013}, their two-parameters parametrization differs from both the Milky Way \citep{Cardelli1989} and \citet{Calzetti2000} attenuation laws, but appear to better describe high-$z$ galaxies. 
However, using the full four-parameter attenuation curve, which covers the \citet{Calzetti2000} curve and other curves from the Milky Way to the Large Magellanic Cloud \citep{Gordon2003}, does not alter my scientific conclusions.

\subsection{Matching photometry and spectra}

I have a spectroscopic redshift for every galaxy from the \hbox{LEGA-C} catalogue. I only included photometric bands in the fit if their FWHM is fully enclosed by the template's wavelength coverage at that redshift. Before the \ppxf\ fits, I pre-computed the $\langle g_n(\lambda)\rangle_q$ and $\lambda_{q,n}$ in  \autoref{eq:model_phot_approx}. I used the same photonic throughput file for all filters, produced for the \textsc{eazy} code \citep{Brammer2008} and available online\footnote{Available from \url{https://github.com/gbrammer/eazy-photoz}}.

Spectroscopy was observed within 1\arcsec\ slits, while photometric observations were either measured within a 2\farcs1 aperture \citep{Muzzin2013} or are total magnitudes \citep{Weaver2022}. This means that calibration is needed to match the flux levels of photometry and spectroscopy. It’s important to note that in some cases, the photometric fluxes may correspond to a different stellar population than that sampled by the spectra. With this in mind, I assumed that spectroscopy and photometry originate from a single spectral energy distribution and applied a constant scaling factor $\kappa$ to the spectrum $G^{\rm spec}(\lambda_p)$. This factor ensures that the synthetic photometry derived from the spectrum using filter transmission curves matches the observed LEGA-C photometry in bands covered by the spectroscopy. This step calibrates the overall normalization of the spectrum flux using photometry before starting the \ppxf\ fit.

To match the VIMOS spectra to the photometry, before calling \ppxf, for every galaxy I first computed synthetic photometric fluxes $\langle f_\lambda \rangle_q$ from its VIMOS spectrum using \autoref{eq:band_integ}, for the subset of photometric bands  (typically 7) contained within the VIMOS wavelength range. I then multiplied the spectrum by a factor $\kappa$ to minimise the $\chi^2$ between the synthetic and observed photometry, only for the few bands in common. It can be computed with the general analytic linear-fitting relation \citep[e.g.][eq.~51]{Cappellari2008}
\begin{equation}\label{eq:scaling}
	\kappa = \frac{\mathbf{d}\cdot\mathbf{m}}{\mathbf{m}\cdot\mathbf{m}},
\end{equation}
where the ``data'' vector $\mathbf{d}$ has elements $d_q=p_q/\Delta p_q$, the observed photometric fluxes $p_q$, divided by their uncertainties $\Delta p_q$ and the ``model'' vector $\mathbf{m}$ has elements $m_q=\langle f_\lambda \rangle_q/\Delta p_q$, the synthetic fluxes also divided by the data uncertainties.

Like \citet{Kriek2013}, I didn't use the catalogues' formal photometric uncertainties in any of my fits, including when calculating $\kappa$ and during \ppxf\ fits. This is because the small error bars at longer wavelengths would have dominated the fits. Additionally, after many fits, it became apparent that systematic imperfections in the SPS model assumptions or data were the main source of uncertainty, rather than random noise. Instead, I use fixed linear uncertainties for all photometric bands of a given galaxy as explained in the next section.

\subsection{Outliers removal}

To remove outliers from the spectral fits, I follow a common practice \citep[e.g.][]{Westfall2019} that involves multiple \ppxf\ fits and adjusting the uncertainties based on the fit residuals. My approach is designed for robustness as follows:
\begin{enumerate}

\item I perform an initial \ppxf\ fit assuming a reasonable fixed uncertainty $\Delta G^{\rm phot}(\lambda_q)$ for all bands of 3\% of the maximum photometric flux $G^{\rm phot}(\lambda_q)$ for that galaxy. Similarly, for the spectrum uncertainty $\Delta G^{\rm spec}(\lambda_p)$, I adopt a constant value of 10\% of the median galaxy spectrum $G^{\rm spec}(\lambda_p)$. A constant uncertainty is a good approximation for the VIMOS spectra and reduces the noise in the fit, with respect to adopting a more accurate but noisy error spectrum (e.g. as given by the reduction pipeline). The best fit is not sensitive to the scaling of uncertainties, which only affects the relative weight of the spectrum and photometry.

\item After the fit, I estimate the rms noise spectrum $\sigma^{\rm spec}_{\rm noise}$ per pixel from the fit residuals, in a statistically robust way, by computing for every spectral pixel the interval containing 68\% of the residuals, within a moving window of 100 pixels.

\item I mask the pixels deviating more than $3\sigma^{\rm spec}_{\rm noise}$ from the best fit. I repeat the masking in a loop, while iteratively adjusting the normalization of the best-fitting spectrum using the non-masked pixels from \autoref{eq:scaling}, until the mask does not change anymore.

\item I multiply $\Delta G^{\rm spec}(\lambda_p)$ by $\sqrt{\chi_{\rm spec}^2/P}$, where $P$ is the number of non-masked spectral pixels, and I compute $\chi_{\rm spec}^2$ from the spectrum alone, in such a way that, after rescaling of the uncertainties, the resulting $\chi_{\rm spec}^2=P$.

\item I do the same constant rescaling for the photometric uncertainties to enforce $\chi_{\rm phot}^2=Q$. Here $Q$ is the number of fitted photometric bands, and I compute $\chi_{\rm phot}^2$ from the photometry alone.

\item After the spectral masking and the rescaling of the photometric $\Delta G^{\rm phot}(\lambda_q)$  and spectral $\Delta G^{\rm spec}(\lambda_p)$ uncertainties, I perform a second \ppxf\ fit, from which I extract the final results.

\end{enumerate}

The rescaled uncertainties are generally of the same order of magnitude as the formal ones provided by the pipelines. However, there can be significant relative differences between different photometric bands. I tested the full LEGA-C sample and found that none of the results in this paper depended on whether I used the formal or rescaled uncertainties. However, the approach I adopted significantly reduced the number of cases where, after visual inspection, the formal best fit did not match the data well because of unrealistically small formal uncertainties that overemphasized certain photometric bands.

\subsection{Gas model and kinematic constraints}

For galaxies without Active Galactic Nuclei, gas emission could be approximately predicted based on the galaxy SFH and could be included in the models, with some extra assumptions, based on photoionization models like \textsc{cloudy} \citep{Ferland1998,Ferland2013}. This feature is implemented in \textsc{fsps} and can be useful when fitting photometry alone where the gas emission are poorly constrained by the data. However, in my case, I have many good-quality spectra in addition to photometry and I want to be able to fit the gas more accurately than a model could predict. For this reason, I fit the gas emission lines in a model-independent way with \ppxf.

With \ppxf\ one can fit many gas emission lines simultaneously to the stellar continuum. This is especially important when studying the stellar population of star-forming galaxies or AGNs, where key absorption lines like the Balmer series are filled by emission. However, when fitting gas lines in relatively low S/N spectra, it is essential to set constraints on the parameters of the gas lines, to prevent possible degenerate situations. An example of a situation to avoid is when the spectrum does not have gas emission and the Gaussian describing an emission line becomes so wide as to become degenerate with the shape of the stellar continuum. 

This is one of the types of situations for which I designed the linearly constrained algorithm of \autoref{sec:capfit}. For my fits to the LEGA-C spectra, after some experimentation focusing on the few problematic fits, I found it sufficient to require the dispersion of the gas emission lines to be smaller than the stellar one $\sigma_{\rm gas}<\sigma_\ast$ and in addition I required the gas and mean stellar velocities to satisfy $|V_{\rm gas}-V_\ast|<500$ \kms. I enforced these requirements as linear constraints in \ppxf\ (keyword \texttt{constr\_kinem}). My strict constraints on the gas dispersion is not always verified in galaxies and I would not recommend it when one is interested in the gas kinematics. However, it appears to work well in eliminating spurious solutions for the stellar population alone from the present type of spectra.

The emission lines that I included in the \ppxf\ fits are all the lines listed in \citet[table~1]{Belfiore2019}. In particular, those falling within the LEGA-C wavelength range for $0.6<z<1$ are the Balmer series bluer than H$\beta$, the [OII]$\lambda\lambda$3726,29, [NeIII]$\lambda\lambda$3868,69 and [OIII]$\lambda\lambda$4959,5007 doublets, and the HeII$\lambda$4687. I force the kinematics of all the gas lines to be the same and I additionally fix the [OIII] doublet to the 1/3 ratio. I fit the Balmer series as a single gas template with decrement for Case B recombination, for temperature $T=10^4$ K and electron density $n_e=100$ cm$^{-3}$ from \citet{Storey1995}. I allow the gas templates to have their own \citet{Calzetti2000} attenuation curve. Fixing the intrinsic ratios of the Balmer series allows me to provide a better extrapolation of the gas filling the weakest (higher-order) absorption lines of the series, even when the S/N of the spectrum is not high enough to constrain them. 

Note that, although I do not include theoretical gas emission predictions in the SPS models, I do include the contribution of the emission lines that are spectroscopically constrained in the photometry. In particular, when Balmer lines are present in the spectrum, the line fluxes of the Balmer series, and in particular of H$\alpha$, which is outside the LEGA-C wavelength range, are included in the photometric fit. However, I checked that this inclusion has a minimal effect on the final results.

\begin{table*}
	\caption{JAM dynamical masses and \ppxf\ stellar population results using SPS templates from \textsc{fsps}.\label{tab:results}}
	
	\begin{tabular}{cccccccccc}
		\hline
		ID LEGA-C &          RA           &          DEC          &  $\lg M_{\rm JAM}$   &    $\lg\Sigma_1^{\rm JAM}$     & $\lg M_*^{\rm ppxf}$ & $\langle\lg {\rm Age}\rangle$ & $\langle[M/H]\rangle$ &     $A_V$      & $\delta$ \\
		          & ($\mathrm{{}^{\circ}}$) & ($\mathrm{{}^{\circ}}$) & ($\mathrm{M_{\odot}}$) & ($\mathrm{M_{\odot}\,kpc^{-2}}$) & ($\mathrm{M_{\odot}}$) &         $(\mathrm{yr}$)         &                       & ($\mathrm{mag}$) &          \\
		   (1)    &          (2)          &          (3)          &         (4)          &              (5)               &         (6)          &              (7)              &          (8)          &      (9)       &   (10)   \\ \hline
		    1     &      150.415222       &       1.758146        &        11.560        &             10.028             &        11.348        &             8.982             &        -0.271         &     1.122      &  0.111   \\
		    4     &      150.412292       &       1.777617        &        11.405        &             9.548              &        11.350        &             9.107             &        -1.040         &     1.356      &  0.377   \\
		    5     &      150.381104       &       1.780024        &        10.847        &             9.436              &        10.947        &             8.719             &        -0.475         &     0.304      &  -0.075  \\
		    6     &      150.449783       &       1.783690        &        11.119        &             9.981              &        11.189        &             9.538             &        -0.140         &     0.000      &  0.400   \\
		    8     &      150.378815       &       1.792945        &        11.653        &             9.919              &        11.445        &             9.088             &        -0.279         &     1.711      &  0.400   \\
		   10     &      150.407440       &       1.803214        &        11.695        &             10.587             &        11.698        &             9.737             &        -0.094         &     0.097      &  0.239   \\
		   11     &      150.423050       &       1.807328        &        11.601        &             9.649              &        11.579        &             9.032             &        -0.102         &     0.075      &  -1.000  \\
		   12     &      150.405838       &       1.813150        &        10.817        &             9.137              &        11.082        &             8.554             &        -0.572         &     0.579      &  -0.165  \\
		   13     &      150.385376       &       1.816353        &        11.113        &             10.036             &        11.190        &             9.285             &        -0.036         &     0.000      &  0.400   \\
		   14     &      150.430817       &       1.820639        &        11.217        &             9.246              &        11.217        &             8.159             &        -0.740         &     0.770      &  -0.038 \\
		   	\hline
	\end{tabular}
	
	{\raggedright {\em Note.} -- Columns (1), (2) and (3): ID, right ascension and declination J2000 in degrees from the LEGA-C catalogue of \citet{vanderWel2021}. Column (4): JAM dynamical masses from \autoref{sec:jam} in solar masses. For accurate quantitative use, one should divide these masses by the luminosities of the Sersic models in the LEGA-C catalogues to obtain the total $(M/L)_{\rm JAM}$ as described in \autoref{sec:dyn_pop}; Column (5): dynamically-determined average mass density within a cylinder of radius $R=1$ kpc along the line-of-sight. I computed this from the best-fitting JAM model;  Column (6): Stellar masses from \ppxf\ using SPS templates from \textsc{fsps}. These masses include living stars and stellar remnants, but exclude gas lost during stellar evolution. I assume a Salpeter IMF with a lower/upper mass cutoff of 0.08 and 100 \msun\ respectively; Columns (7) and (8): $\lg {\rm Age}$ and $[M/H]$ weighted by the bolometric luminosity (\autoref{sec:stellar_pop_scaling}). Columns (9) and (10): $V$-band attenuation in mag and slope $\delta$ from \autoref{eq:attenuation}, for the two-parameters attenuation described in \autoref{sec:dust}. I show only the first ten rows of this table, while the full electronic table for 3197 galaxies (including duplicates) is available as Supporting Information from the MNRAS website.
		\par}
\end{table*}

\begin{figure*}
	\includegraphics[width=\textwidth]{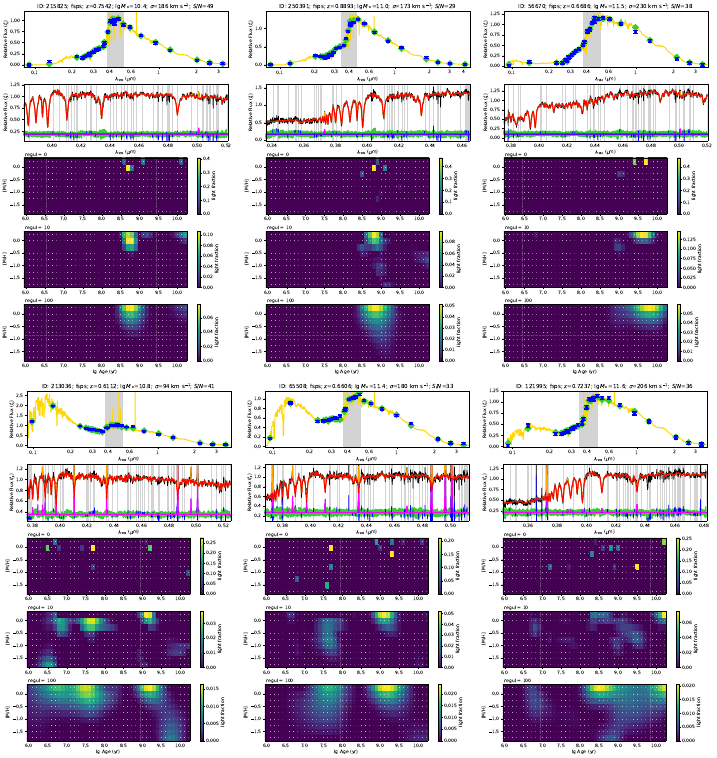}		
	\caption{Examples of \ppxf\ fits to the LEGA-C galaxy spectra and 28-bands photometry using the \textsc{fsps} models. The top three galaxies require a model with a single short star formation event, while the bottom three galaxies require multiple discrete star formation events. For each galaxy, the top panel shows the photometric measurements (blue error bars) and the best fit (green diamonds), while the golden line shows the underlying best-fitting template with included emission lines. The grey vertical band indicates the range where spectroscopy was also fitted. The second panel shows the observed spectrum (black line) and the best-fitting total spectrum (orange line). The best-fitting stellar spectrum alone is shown in red and the gas emission one is in magenta. The residuals (arbitrarily offset) are indicated with green diamonds and the masked pixels with blue lines (and corresponding grey vertical bands). The last three panels show the distribution of the \ppxf\ weights, indicating the bolometric luminosity $L_{\rm bol}$ of each stellar population of given age and metallicity. The weights are shown for (i) no regularization, (ii) regularization \texttt{regul=10} and (iii) \texttt{regul=100} as written in the plots.
		\label{fig:ppxf_vary_regul}}
\end{figure*}

\begin{figure*}
	\includegraphics[width=\textwidth]{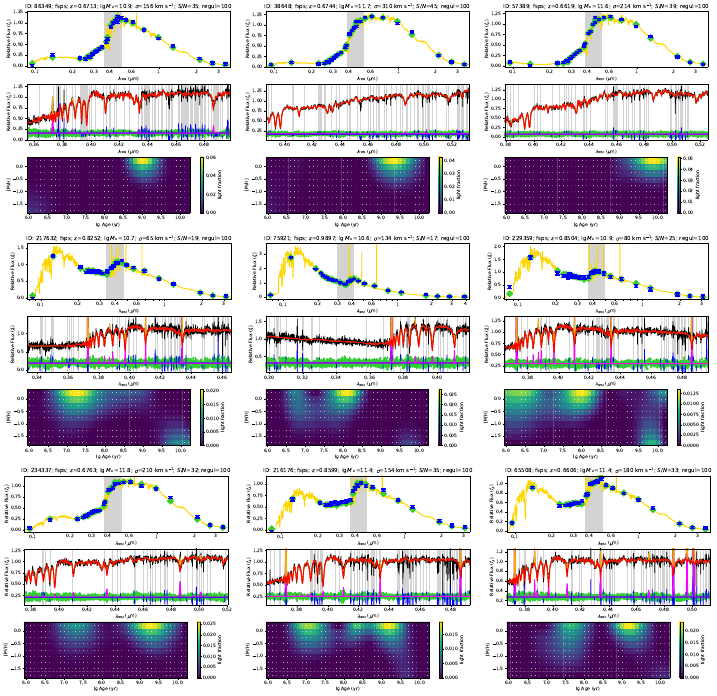}		
	\caption{More examples of \ppxf\ fits to spectra and photometry. The meaning of the symbols is the same as in \autoref{fig:ppxf_vary_regul}, but I show a single regularization (\texttt{regul=100}). The top three galaxies require a single star formation event, while the rest can only be modelled with multiple discrete star formation events. \label{fig:ppxf_regul100}}
\end{figure*}

\subsection{Velocity dispersion matching}

The LEGA-C data have an instrumental dispersion of $\sigma_{\rm inst}\approx36$ \kms\ (\autoref{sec:spectroscopy}), ignoring possible variations within the rather small wavelength range. The MILES stellar templates used in both the \textsc{fsps} and \textsc{galaxev} models were observed with an instrumental resolution of $\Delta\lambda\approx2.50$ \AA\ FWHM \citep{FalconBarroso2011miles}, equivalent to $R=\lambda/\Delta\lambda\approx1600$ at the typical wavelength $\lambda\approx0.4$ \micron\ covered by LEGA-C. This corresponds to an instrumental dispersion
\begin{equation}
	\sigma_{\rm inst}=\frac{c}{R\,\sqrt{4\ln4}}\approx80\,\kms.
\end{equation}

Ideally, I would like to use SPS models based on stars with higher resolution than the galaxy spectra. However, assuming that the instrumental line spread functions are approximately Gaussian, one can still use a template with higher instrumental dispersion $\sigma_{\rm inst,tem}$ than that $\sigma_{\rm inst,gal}$ of the observed galaxy spectrum as long as $(\sigma_{\rm inst,gal}^2 + \sigma_\ast^2) > \sigma_{\rm inst,tem}^2$, where $\sigma_\ast$ is the real ``astrophysical'' dispersion of the galaxy stars. After the \ppxf\ fit one can compute the corrected stellar dispersion with the standard expressions \citep[sec.~2.2]{Cappellari2017}
\begin{subequations}
	\begin{align}
		&\sigma_{\rm diff}^2 = \sigma_{\rm inst,gal}^2 - \sigma_{\rm inst,tem}^2\\
		&\sigma_\ast^2 = \sigma_\ppxf^2 - \sigma_{\rm diff}^2.
	\end{align}
\end{subequations}
As $\sigma_{\rm diff}^2$ is in this case a negative quantity, I can model the dispersion of galaxies down to $\sigma_\ast\ga|\sigma_{\rm diff}^2|^{1/2}=71$ \kms. I compared my fitted dispersions $\sigma_\ast$ with the values in the LEGA-C DR3 catalogue, which were measured with \ppxf\ using higher resolution synthetic templates as described in \citet{Bezanson2018}. I found a good agreement assuming $|\sigma_{\rm diff}^2|^{1/2}\approx48$ \kms, which suggests possible inaccuracies in the quoted relative instrumental dispersion of the galaxies and the templates. Regardless of the reason for this discrepancy, only 40 of the 3197 galaxies in the catalogue with measured dispersion have $\sigma_\ast<48$ \kms, likely due to measurement uncertainties. This implies that I can safely use the SPS based on MILES models to study the stellar population of LEGA-C galaxies.

\section{Results}
\label{sec:results}

In this section, I describe the results of my stellar population modelling with \ppxf. I also compare masses from stellar population and galaxy dynamics. The key quantities used in this paper are given in \autoref{tab:results}.

\subsection{Spectral fit examples}

In this paper, I focus on galaxy observable trends rather than on comparisons with models of galaxy formation. For this reason, instead of converting the SFH recovered by \ppxf\ into stellar masses formed in a given time interval, I will always show the fraction of bolometric luminosity contributed by each template, as a function of their age and metallicity $[M/H]$. More precisely, I integrate the luminosity from the template spectra only within the region $0.1<\lambda<3$ \micron\ covered by the data. This is to avoid the possibility of interpreting very young stars, which emit most of their luminosity for $\lambda<0.1$ \micron, as contributing significantly to my observables, even when their flux is not detected in the data, but simply extrapolated. I still indicate my luminosity as $L_{\rm bol}$ because, except for extremely young stars, it still represents a very good approximation for it. 

The advantage of using $L_{\rm bol}$ rather than SFH, is that one can get a direct sense of what the data actually show, without strongly nonlinear conversions into masses, due to the large $M/L$ differences of different stellar populations. In fact, I would argue that comparisons with models of galaxy formations are generally more meaningful when the models, for which all quantities are known accurately, are converted into luminous observables, rather than trying to do the reverse by extracting SFH in masses from the data.

In the course of this study, I fitted the 3197 galaxies of my subsample (\autoref{sec:spectroscopy}) with \ppxf\ multiple times with different levels of regularization, or no regularization at all, to test the sensitivity of the results. In \autoref{fig:ppxf_vary_regul} I illustrate the effect of regularization on some high-$S/N$ spectra. These figures, like \autoref{fig:montecarlo}, illustrate the ill-conditioning of the stellar population inversion, which prevents one from obtaining a unique solution, even from very good data. Nonetheless, the figure also illustrates the ability of the method to distinguish the striking difference between (i) galaxies that can only be described, even at high regularization\footnote{A given value of the \ppxf\ keyword \texttt{regul} roughly implies that neighbouring weights $w_{ij}$ can differ by $\Delta w_{ij}\sim1/\texttt{regul}$. As I normalize all galaxy spectra to the same average flux (e.g. average$=$1), setting a given \texttt{regul} value roughly corresponds to requiring a similar level of smoothness in the distribution of the weight.} (\texttt{regul=100}) by a single star formation event at a very localized $\lg{\rm Age}$ (top three panels in \autoref{fig:ppxf_vary_regul}) and (ii) galaxies that require multiple and separated star formation events to be described (bottom three panels in \autoref{fig:ppxf_vary_regul}). The galaxies in the top panels are essentially described by a single SPS model, from 0.1 \micron\ to 3 \micron, for both spectra and photometry. This highlights the success of the SPS models in accurately predicting real galaxy spectra.

In \autoref{fig:ppxf_regul100} I show additional examples of \ppxf\ fits to good quality spectra to give a sense of the variety of spectral morphologies and the corresponding variations in the $L_{\rm bol}$ weights distributions. I used in all these cases a high regularization (\texttt{regul=100}). Also here one can clearly see the striking difference between (i) the three galaxies in the top row, which can only be described as a single burst of star formation, which happened at different times and (ii) galaxies requiring multiple discrete star formation events. Star formation events appear to have a similar extent in $\ln {\rm Age}$, which seems to imply that events in the past lasted longer than recent ones. This is likely an artefact of our general ability to more accurately detect age differences in recent events.

\subsection{Comparing \ppxf\ stellar masses with other methods}

\begin{figure} % trim={left bottom right top}
	\centering
	\includegraphics[width=0.9\columnwidth]{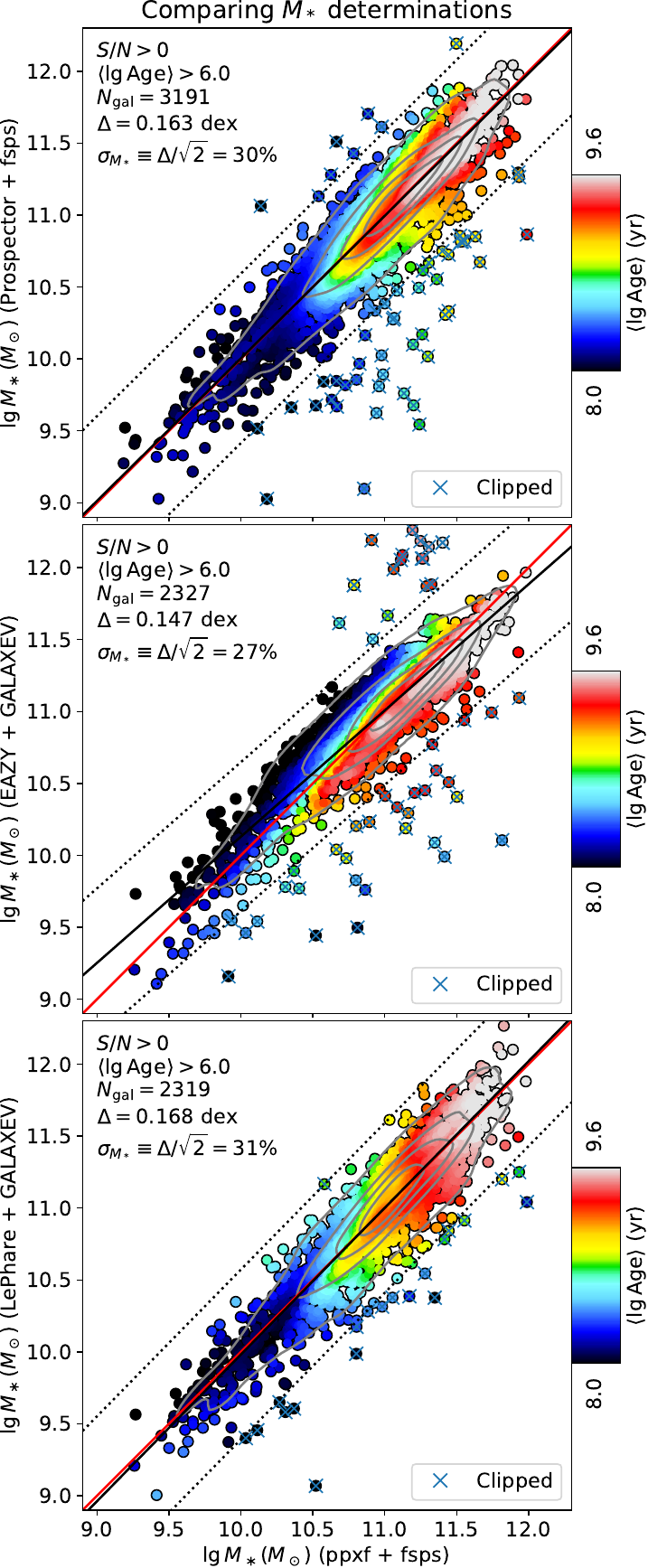}					\caption{Comparison of \ppxf\ stellar population mass $M_*$ estimates against different methods. I use the $M_*$ values from \textsc{Prospector} in \citet{vanderWel2021} or from \textsc{EAZY} and \textsc{LePhare} in \citet{Weaver2022}, for the galaxies that are common in both catalogues. The three panels show the comparison for each method. The red line is the one-to-one relation, and the black line is the best fit using \textsc{LtsFit} \citep{Cappellari2013p15}. The x-symbols are the outliers identified by \textsc{LtsFit}, and the dotted lines are the selection limits. The colours show the luminosity-weighted mean stellar age $\langle\lg {\rm Age}\rangle$, smoothed by \textsc{loess}. The grey contours show the kernel density estimation of the galaxy distribution. I indicate the $S/N$ and $\langle\lg {\rm Age}\rangle$ selection criteria, the number of selected galaxies, the rms scatter $\Delta$, and an approximate relative $1\sigma$ error $\sigma_{M_*}$ in $M_*$ for each panel. I estimated the latter as $\sigma_{M_*}\equiv\Delta/\sqrt{2}$ by assuming both $M_*$ have the same uncertainty.  \label{fig:stellar_mass_accuracy}}
\end{figure}

\begin{figure} % trim={left bottom right top}
	\centering
	\includegraphics[width=0.9\columnwidth]{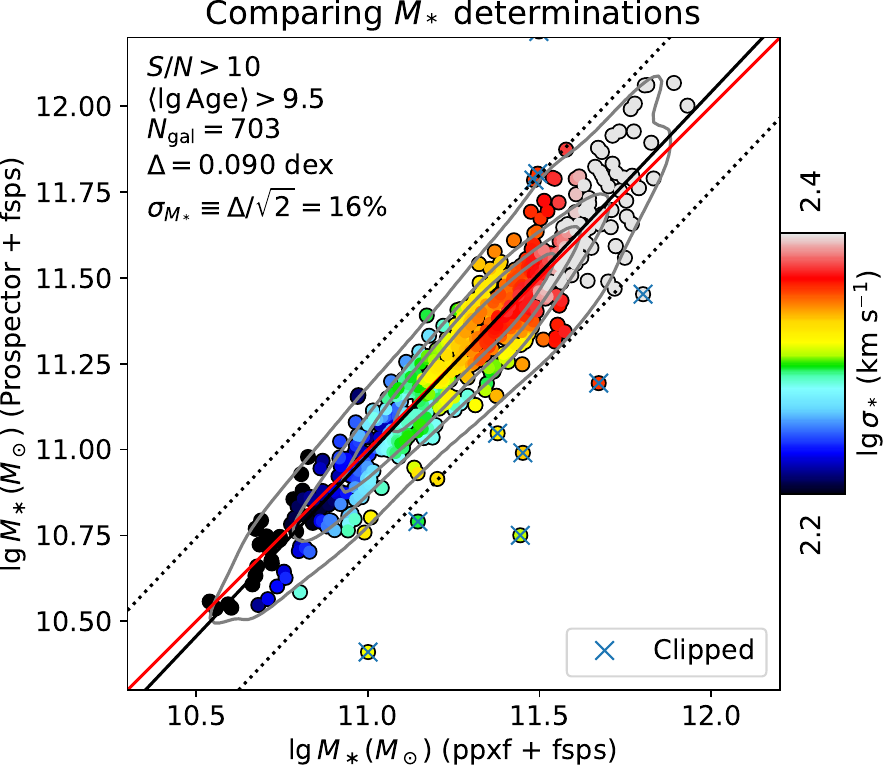}
	\caption{Same as in \autoref{fig:stellar_mass_accuracy} for a subsample with high spectral $S/N>10$ and old ages $\langle\lg {\rm Age}\rangle>9.5$. As the age range is limited, the colours show here the \textsc{loess} smoothed stellar velocity dispersion. \label{fig:stellar_mass_old_accuracy}}
\end{figure}

The weights for different ages and metallicities produced by a fit with \ppxf\ can be converted into stellar masses. In this section I compare the masses derived with \ppxf\ against the stellar masses produced by other codes.

For my comparisons I used the published stellar masses $M_*$ for the LEGA-C galaxies from the three stellar population codes: (i) \textsc{LePhare} \citep{Arnouts2002}, (ii) \textsc{EAZY} \citep{Brammer2008} and (iii) \textsc{Prospector} \citep{Johnson2021}. I extracted the values of stellar masses for the first two codes from the \textsc{Farmer} version of the COSMOS2020 catalogue \citep{Weaver2022}, while for the last code, I used the values described in the LEGA-C DR3 paper \citep[appendix~B]{vanderWel2021} as kindly provided by Arjen van der Wel.

I tried to isolate the effect of the fitting methods from differences in the extrapolation of the galaxy's total luminosities. For this, with the $M_*$ from the COSMOS2020 catalogues, I rescaled the masses by the difference in the total $K_s$ band luminosity between the \citet{Weaver2022} and \citet{Muzzin2013} catalogues. No correction is needed to compare \ppxf\ and \textsc{Prospector}, given that for both I used the values based on the \citet{Muzzin2013} catalogue. 
 
I show the comparison between the stellar masses derived with \ppxf\ and the other three codes in \autoref{fig:stellar_mass_accuracy}. I coloured the values with the stellar age derived by \ppxf. I \textsc{loess} smoothed the measured age values using the algorithm by \citet{cleveland1988locally} as implemented in the \textsc{loess} package\footnote{I used \textsc{loess} v2.1 available from \url{https://pypi.org/project/loess/}} by \citet{Cappellari2013p20} and using the keyword \texttt{rescale=True} to equalize the axes of maximum/minimum variance before smoothing. I used a small smoothing parameter \texttt{frac=0.1} in all plots of this paper. The \textsc{loess}-smoothed values are the two-dimensional equivalent of the average trend that is often shown in one-dimensional plots. The key difference is that the scatter cannot be easily shown in two dimensions together with the average trend. The scatter is better visualized using a different projection.

To estimate the scatter between two pair of measurements, while removing outliers, I used the \textsc{LtsFit} package\footnote{I used \textsc{LtsFit} v6.0 available from \url{https://pypi.org/project/ltsfit/}} 
described in \citet[sec.~3.2]{Cappellari2013p15}, which combines the Least Trimmed Squares robust technique of \citet{Rousseeuw2006} into a least-squares fitting algorithm which allows for errors in all variables and intrinsic scatter. Instead of using a fixed $\sigma$-clipping criterion with the \texttt{'clip'} keyword in the \texttt{ltsfit} procedure, I used an adaptive clipping that depends on the sample size. This is the value that would produce on average one outlier in a Gaussian distribution of the given sample size. It can be computed using the Scipy class \href{https://docs.scipy.org/doc/scipy/reference/generated/scipy.stats.norm.html}{\texttt{scipy.stats.norm}} as \texttt{clip=abs(norm.ppf(p/2))}, with $p=1/n$ and $n$ the sample size. For reference, with $n=100$ this gives \texttt{clip=2.58} (default for \texttt{ltsfit}), for $n=500$, \texttt{clip=3.09} and for $n=3000$, \texttt{clip=3.59}. The \texttt{ltsfit} procedure returns a robust estimate of the rms scatter $\Delta$ from the best-fitting relation. When the uncertainty of the two quantities I am comparing is the same, one can estimate it as $\sigma_{M_*}=\Delta/\sqrt{2}$.

In all my plots I rescaled the masses provided by all other methods to have the same median as the \ppxf\ values, which I did not modify. This is the reason why all plots follow the one-to-one relation without any overall offset. This is to remove the effect of differences in the assumed stellar IMF, gas loss or stellar remnants, whose investigation is outside the scope of this paper. I find that the observed scatter in all galaxies, when selected irrespective of their age or $S/N$, is in agreement with a $1\sigma$ uncertainty in the stellar mass of about 30\% for every method. This result is consistent across all six pairwise comparisons of the methods, with differences within the measurement uncertainties. However, the behaviour of the differences is markedly different as a function of mean ages. The comparison of \ppxf\ against \textsc{Prospector} (\autoref{fig:stellar_mass_accuracy}), show that the scatter is smaller for older galaxies at given mass, but the younger ones generally scatter symmetrically around the one-to-one relation. The exception are the outliers, which have generally lower masses in \textsc{Prospector} than in \ppxf. The comparison of \ppxf\ and \textsc{LePhare} is similar to \textsc{Prospector}, but with less low-mass outliers. However there is no evidence for a tightening of the correlation for older models. The comparison of \ppxf\ and \textsc{EAZY}, unlike the other two models, shows a strong asymmetry as a function of age: older models tend to be less massive in \textsc{EAZY} than \ppxf, while younger models are more massive in \textsc{EAZY}. This asymmetry is reminiscent of the difference between \textsc{Prospector} and \textsc{EAZY} reported in \citet{Leja2019}. In fact, the same age asymmetry is seen when comparing \textsc{EAZY} with either \textsc{Prospector} or \textsc{LePhare}.

As suggested by \autoref{fig:stellar_mass_accuracy}, the scatter dramatically decreases (\autoref{fig:stellar_mass_old_accuracy}) if I compare \ppxf\ and \textsc{Prospector} only for the galaxies with the oldest ages and largest spectral $S/N$ (which also implies brightest photometry). Given the small age range, I coloured galaxies by their $\sigma_*$. For this subset of galaxies, the inferred scatter of about 16\% is half of that for the general population, without significant trends, except again for some outliers where \textsc{Prospector} gives lower masses than \ppxf.

When comparing stellar mass estimates of real galaxies, it is often difficult to assess the real accuracy between different methods, because the true masses are unknown. In the next section, I will address this issue, for a subsample of the LEGA-C sample, using mass determinations from stellar dynamics.

\subsection{Comparing JAM dynamical with stellar population $M/L$}
\label{sec:dyn_pop}

\begin{figure} % trim={left bottom right top}
	\centering
	\includegraphics[width=0.9\columnwidth]{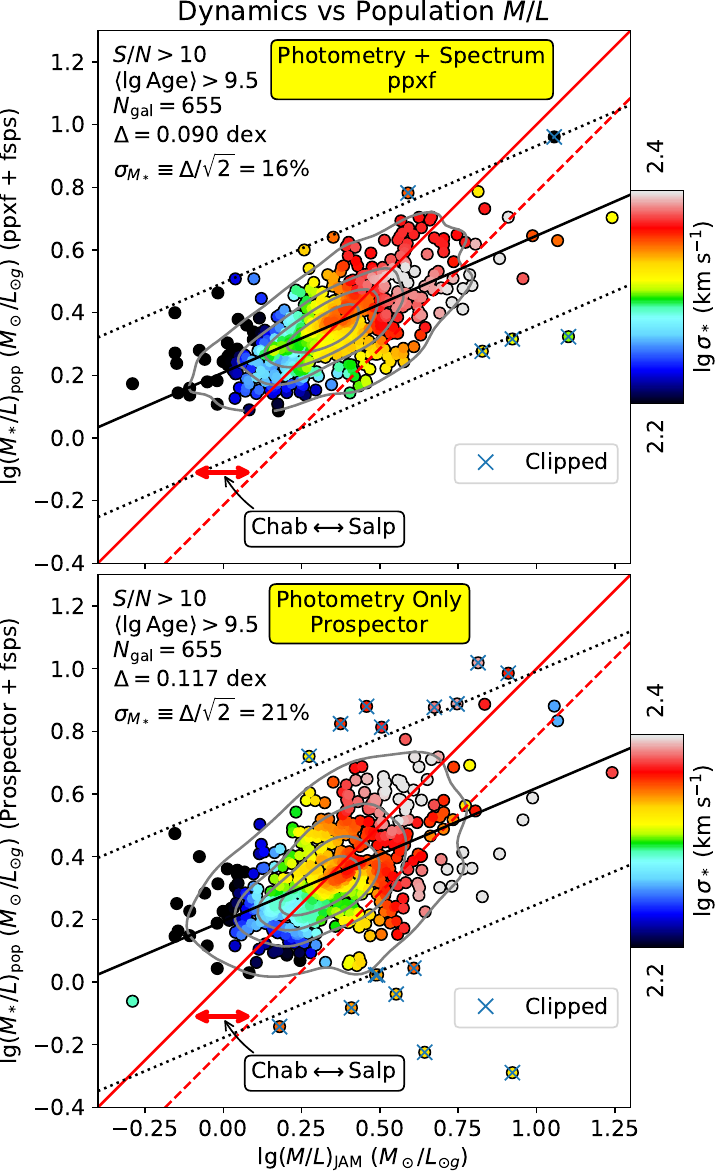}
	\caption{Comparison of dynamical and stellar population mass-to-light ratios. The dynamical mass-to-light ratio $(M/L)_{\rm JAM}$ is derived from the JAM modelling, while the stellar population mass-to-light ratio $(M/L)_{\rm pop}$ is derived from either \ppxf (top panel) or \textsc{Prospector} (bottom panel). The red line shows the one-to-one relation, while the red dashed line indicates the shift that would be produced by changing the IMF from Chabrier to Salpeter. The black line shows the best linear fit using \textsc{LtsFit}, which strongly deviates from the one-to-one relation. The x-symbols mark the outliers identified by \textsc{LtsFit}, and the dotted lines mark the selection limits. The colours indicate the stellar velocity dispersion $\sigma_*$, smoothed by \textsc{loess}. The grey contours indicate a kernel density estimate of galaxies. For each panel, I show the signal-to-noise ratio $S/N$ and the mean logarithmic age $\langle\lg {\rm Age}\rangle$ selection criteria, the number of selected galaxies $N$, the root mean square scatter $\Delta$, and an approximate relative error $\sigma_{M_*}$ in stellar mass $M_*$ for both methods. I estimate $\sigma_{M_*}$ as $\sigma_{M_*}\equiv\Delta/\sqrt{2}$ by assuming both methods have the same uncertainty.  \label{fig:jam_vs_pop_ml_accuracy}}
\end{figure}

\begin{figure} % trim={left bottom right top}
	\centering
	\includegraphics[width=0.9\columnwidth]{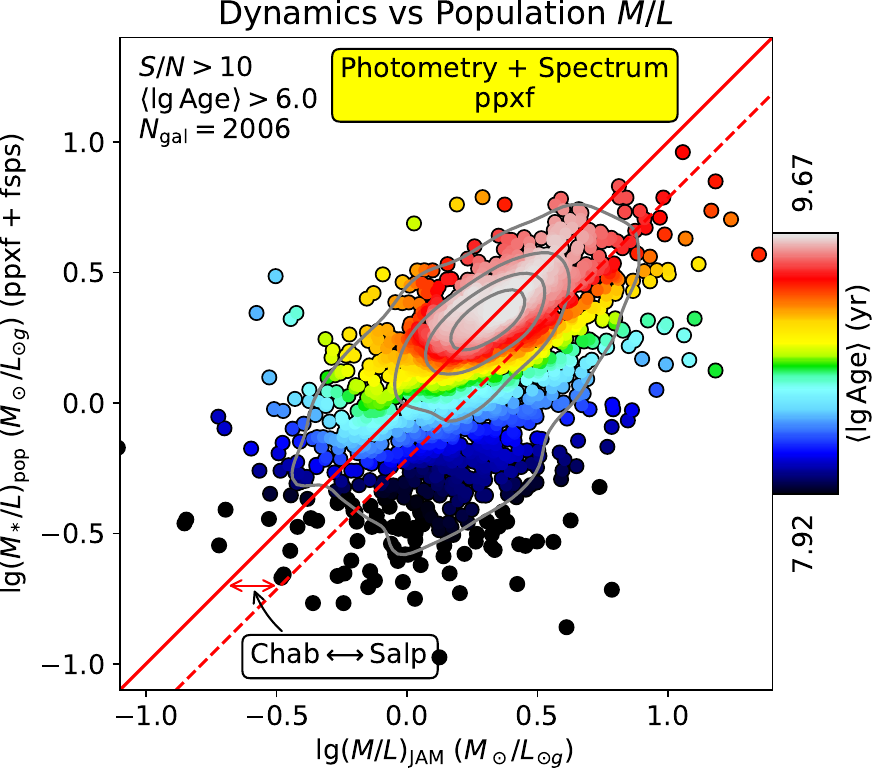}
	\caption{Same as in \autoref{fig:jam_vs_pop_ml_accuracy} but for all ages. The colours here indicate \textsc{loess}-smoothed mean luminosity-weighted ages $\langle\lg {\rm Age}\rangle$. The contours of constant age are almost horizontal, meaning that the stellar population mass-to-light ratio $(M/L)_{\rm pop}$ depends on age as expected, but the dynamical  $(M/L)_{\rm JAM}$ does not. Variations in the initial mass function (IMF) can explain the $(M/L)_{\rm JAM}$ differences for the old ages, while the mass fraction of dark matter and gas must drive the $(M/L)_{\rm JAM}$ variations for the young ages. \label{fig:jam_vs_pop_ml_all_ages}}
\end{figure}

One of the sources of confusion in comparing galaxy masses from stellar populations and dynamical modelling is the ambiguity of the so-called `dynamical mass' of a galaxy. This term does not refer to a well-defined physical quantity, because in the standard cosmological model, the galaxy's total mass is largely composed of dark matter, which is difficult to constrain with the available kinematic data of limited radial coverage. The quantity that the dynamical models reliably measure is the total density profile within the spatial extent of the kinematic tracer. However, this density profile cannot be easily converted into a mass, because it depends on the choice of the integration volume and on the assumptions about the galaxy shape and orientation \citep[sec.~3.3.1]{Cappellari2013p15}. A more robust and convenient quantity for comparing population and dynamics is the total mass-to-light ratio $(M/L)_{\rm JAM}$, within the inner regions of a galaxy. This quantity has a weak dependence on the integration volume and galaxy inclination. It should be always preferred for accurate comparisons.

In \autoref{sec:jam_of_sersic}, I presented unbiased dynamical models of the stellar kinematics, based on the Sersic photometric models in the F814W/ACS band, for all the galaxies in the LEGA-C sample with available velocity dispersion. Previous studies of nearby galaxies using high-resolution integral-field stellar kinematics have demonstrated that this kind of models can reliably estimate the total dynamical mass-to-light ratios $(M/L)_{\rm JAM}$ in the central regions of galaxies with uncertainties of about 5\% \citep{Cappellari2006,Cappellari2013p15,Shetty2020,Zhu2023a}. Importantly, these studies have also shown that precise and unbiased $(M/L)_{\rm JAM}$ can be obtained using models where the total mass distribution follows the luminous one. In fact, these mass-follow-light models are more robust and precise than those that explicitly separate the luminous and dark matter, when the main goal is to measure the total $(M/L)_{\rm JAM}$ \citep{Cappellari2013p15,Zhu2023}.

An additional complication is that the dynamics is sensitive to all mass components: stellar, gas, and dark matter, while the population only measures the stellar one. However, detailed nearby studies have shown that for the passive galaxy population, dark matter contributes only about $\approx10\%$ of the total mass within $1R_{\rm e}$ \citep{Cappellari2013p15,Zhu2023}, while gas mass has an even smaller contribution \citep[e.g.][]{Young2011}. This allows us to assume that the dynamical $(M/L)_{\rm JAM}$ accurately approximates the stellar one.

To compare the dynamical $(M/L)_{\rm JAM}$ I need the same quantity from stellar population $(M/L)_{\rm pop}$. Having the full spectra from the stellar population models, one can compute the $(M/L)_{\rm pop}$ in any band. However, since I only have total stellar masses $M_*$ from \textsc{Prospector} LEGA-C catalogue, I divide $M_*$ by the total luminosity $L_i$ in SUBARU $i$-filter from \citet{Muzzin2013} catalogues. This assumes that $(M/L)_{\rm pop}$ is constant over the full galaxy, which is likely a decent approximation for passive galaxies. Using photometry consistent with mass derivation ensures no spurious differences in mass and luminosity extrapolation. However, differences between $i$-band and F814W filters may introduce some small systematic offset in the $M/L$ comparison. However, I am interested in relative uncertainties more than absolute offsets. The $M/L$ is usually reported in solar units. For this, I assume a solar luminosity $L_g=5.11$ mag in AB system from \citet{Willmer2018} and report $M/L$ in units of $\msun/\lsun$, given that F814W approximately corresponds to rest-frame SDSS $g$-band filter at median redshift $z\approx0.8$ of my sample. This normalization is a constant and does not affect the comparison. The rest-frame wavelength of the filter varies by up to $\approx10\%$ within the redshift range, however, this shift is the same for both dynamical and population $M/L$ and does not affect the scatter.

I compare the $(M/L)_{\rm JAM}$ from dynamical modelling and the $(M/L)_{\rm pop}$ from stellar population synthesis using a sample of old galaxies with high-quality spectra in \autoref{fig:jam_vs_pop_ml_accuracy}. I only use the passive population for this comparison. I adjust the \textsc{Prospector} $(M/L)_{\rm pop}$ by adding 0.19 dex to match the JAM median. I also convert the \ppxf\ $(M/L)_{\rm pop}$ from the \citet{Salpeter1955} IMF to the \citet{Chabrier2003} IMF by subtracting 0.215 dex \citep[fig.~4]{Madau2014}. I do not change the \ppxf\ value after this conversion.

The main findings from \autoref{fig:jam_vs_pop_ml_accuracy} are: 
\begin{enumerate}
	\item The \ppxf\ $(M/L)_{\rm pop}$ values are more consistent with the $(M/L)_{\rm JAM}$ values than the \textsc{Prospector} ones. The scatter is 0.090 dex for \ppxf\ and 0.117 dex for \textsc{Prospector}. This suggests that adding spectra to \ppxf\ improves the mass estimates.
	\item The \ppxf\ $(M/L)_{\rm pop}$ comparison does not have the low-$M/L$ outliers that appear in the \textsc{Prospector} comparison, indicating more reliable $(M/L)_{\rm pop}$ or $M_*$ estimates in \ppxf\ with spectra than in \textsc{Prospector} with photometry only.
	\item Both \ppxf\ and \textsc{Prospector} show a similar trend in the $(M/L)_{\rm pop}-(M/L)_{\rm JAM}$ relation, which clearly deviates from a one-to-one relation. The trend implies that the galaxies with higher $\sigma_*$ have more mass from dynamics than from population models at a fixed IMF. The variation is comparable to the mass difference between Chabrier and Salpeter IMF. This trend is consistent with previous studies that suggested a non-universal IMF based on dynamics and population of nearby \citep{Cappellari2012,Li2017imf,Shetty2020a} and distant galaxies \citep{Shetty2014}. Whatever the origin of this trend, this comparison shows that it is robust across different samples, redshift and methods.
\end{enumerate}

From the cross-comparisons between the scatter observed when comparing different estimates of the stellar masses, one can infer the accuracy of each individual technique, assuming as an approximation that it is constant. In fact, if we define $\sigma_{\rm method}$ the uncertainty of `method', then the squared uncertainties between each pair of methods add linearly as follows
\begin{equation}
\left\{ 
\begin{array}{l}\label{eq:uncertainties}
	\sigma_{\ppxf}^2 +  \sigma_{\rm Pros}^2 = \Delta^2(\ppxf/{\rm Pros}) \\
	\sigma_{\ppxf}^2 +  \sigma_{\rm JAM}^2 = \Delta^2(\ppxf/{\rm JAM}) \\
	\sigma_{\rm Pros}^2 +  \sigma_{\rm JAM}^2 = \Delta^2({\rm Pros}/{\rm JAM})
\end{array} 
\right.
\end{equation}
where the scatter $\Delta$ was measured in \autoref{fig:stellar_mass_old_accuracy} ($\Delta(\ppxf/{\rm Pros})=0.090$ dex) and \autoref{fig:jam_vs_pop_ml_accuracy} ($\Delta(\ppxf/{\rm JAM})=0.090$ and $\Delta({\rm Pros}/{\rm JAM})=0.117$ dex). The positive solution of \autoref{eq:uncertainties} gives the $1\sigma$ relative uncertainty of the three different methods on this dataset: 
\begin{equation}
	\left\{ 
	\begin{array}{l}
		\sigma_{\ppxf} = 0.035\, {\rm dex} = 9\, \%\\
		\sigma_{\rm JAM} = 0.083\, {\rm dex} = 21\, \%  \\
		\sigma_{\rm Pros} = 0.083\, {\rm dex} = 21\, \%
	\end{array} 
	\right.
\end{equation}

This result shows that, at least for the limited case of the old population, where we can assume we know the `true' stellar mass from galaxy dynamics, the inclusion of spectra in \ppxf\ gives masses significantly more accurate than those using \textsc{Prospector} with photometry alone. This is encouraging, but of course, it should not be interpreted as \ppxf\ being more accurate than \textsc{Prospector}, given that the latter could fit spectra as well and this would likely lead to comparable accuracy as \ppxf. However, these extra comparisons are beyond the scope of this paper.

In \autoref{fig:jam_vs_pop_ml_all_ages} I also show the comparison between stellar dynamics and stellar population $M/L$ for the full set of galaxies with high-$S/N$ regardless of their age. This plot cannot be used to infer the accuracy of the mass estimates. In fact, detailed modelling of the MaNGA survey has shown that younger galaxies contain significant fractions of gas and dark matter \citep{Zhu2023a}, making the mass estimate from the stellar population significantly lower than the dynamical one, as observed.

\subsection{Stellar population scaling relations}
\label{sec:stellar_pop_scaling}

\begin{figure*}
	\includegraphics[width=\textwidth]{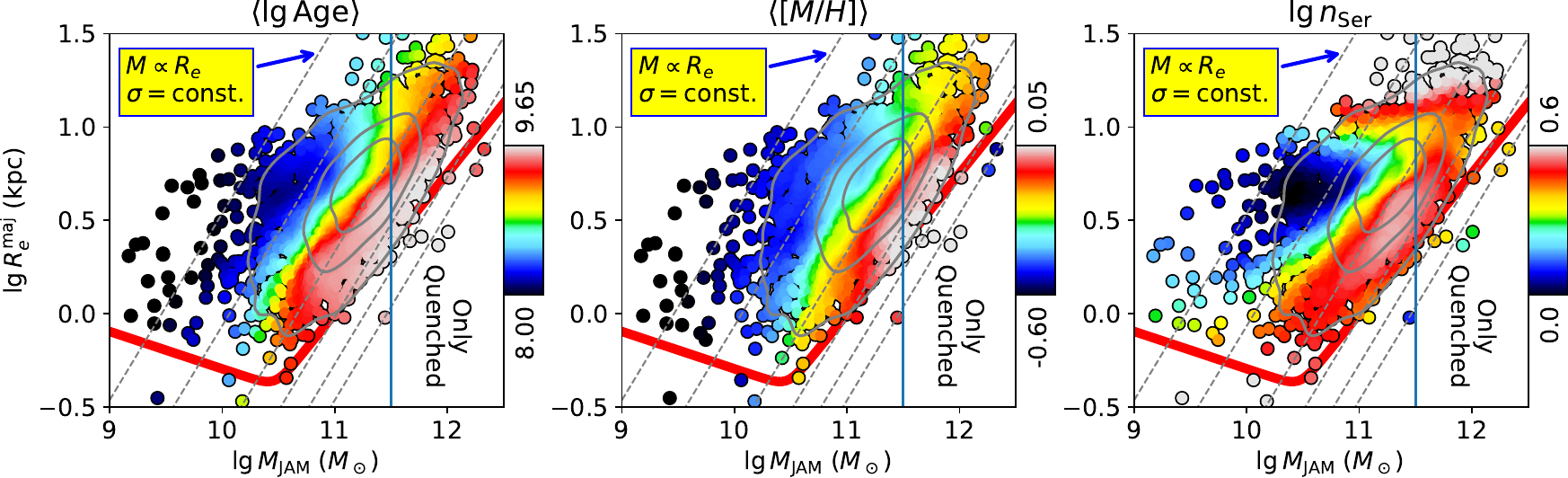}
	\caption{Galaxy properties on the dynamical mass vs size $(M_{\rm JAM},R_{\rm e}^{\rm maj})$ plane at $z\approx0.8$. To good approximation $M_{\rm JAM}\approx M_*$, see text. Galaxies are coloured by their luminosity-weighted ages $\langle\lg {\rm Age}\rangle$ (left panel), luminosity-weighted metallicities $\langle[M/H]\rangle$ (middle panel) and by the Sersic index $n_{\rm Ser}$ of a fit to their photometry. I used the values from the \textsc{fsps} models, but those from \textsc{galaxev} are very similar. All values are \textsc{loess}-smoothed to show average trends and I used \texttt{regul=10} in \ppxf. The grey contours are a kernel density estimate of the galaxy distribution (using \href{https://docs.scipy.org/doc/scipy/reference/generated/scipy.stats.gaussian_kde.html}{\texttt{scipy.stats.gaussian\_kde}}). Galaxy properties mainly follow lines of constant stellar velocity dispersion, which is indicated by the dashed grey lines for $\sigma_\ast=50,100,200,300,400,500$ \kms\ from left to right, while mass is not a good predictor of their stellar population. However, above a stellar mass $\lg (M_\ast/\msun)\ga11.5$ (blue vertical line), all galaxies are old (quenched), have a high metallicity and large Sersic index. The thick red line is the ``zone of avoidance'' for nearby galaxies from \citet{Cappellari2013p20}, scaled down by a factor $1.6\times$ to account for redshift evolution.
		\label{fig:mass_size}}
\end{figure*}

\begin{figure*}
	\includegraphics[width=\textwidth]{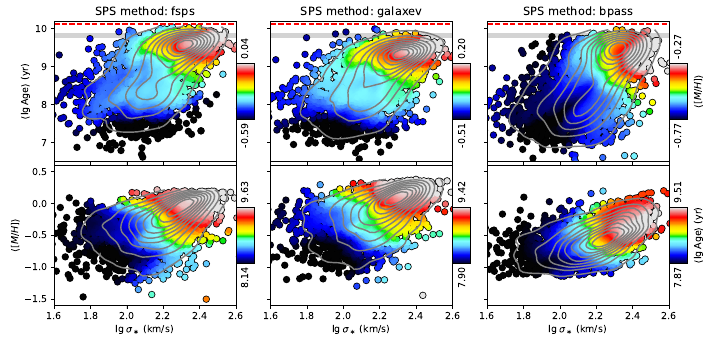}
	\caption{Luminosity-weighted ages $\langle \lg {\rm Age} \rangle$ and metallicities $\langle [M/H] \rangle$ vs stellar velocity dispersion $\sigma_\ast$ for the LEGA-C galaxies. The top panels show the age distribution, coloured by \textsc{loess}-smoothed metallicity. The grey horizontal band is the Universe age range between $0.6<z<1$ and the red dashed line is the present age. The bottom panels show the metallicity distribution coloured by \textsc{loess}-smoothed age. In all panels, the grey contours are the kernel density estimator of the galaxies' distribution. From left to right I show results using the \textsc{fsps}, \textsc{galaxev} and \textsc{bpass} SPS models, all with \texttt{regul=10} in \ppxf. There is a clear bend of the $({\rm Age},\sigma_\ast)$ trend around $\lg(\sigma_\ast/\kms)\approx2.3$. At fixed $\sigma_\ast$ metallicity depends on age with younger galaxies having lower $\langle [M/H] \rangle$. 
		\label{fig:sig_age_metal}}
\end{figure*}

\begin{figure*}
	\includegraphics[width=\textwidth]{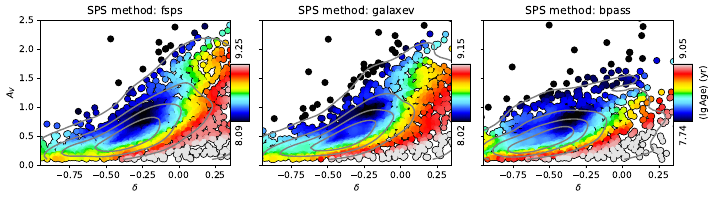}
	\caption{Dust properties vs galaxy age. The $V$ band attenuation $A_V$ in mag is plotted against the UV slope $\delta$ and coloured by \textsc{loess}-smoothed luminosity-weighted ages. The grey contours are the kernel density estimator of the galaxies' distribution. At every $\delta$ the youngest galaxies have larger attenuation, while the largest $A_V$ are only observed around the \citet{Calzetti2000} attenuation curve, which has $\delta=0$. From left to right I show results using the \textsc{fsps}, \textsc{galaxev} and \textsc{bpass} SPS models. The first two are highly consistent, but the latter, while still qualitatively similar is quantitatively very different. \label{fig:dust}}
\end{figure*}

As I am focusing on observable trends, I define luminosity-weighted population quantities, summed over the template weights, as
\begin{subequations}
	\begin{align}
		&\langle \lg {\rm Age}\rangle = \frac{\sum_{j} L_{{\rm bol},j}\times\lg {\rm Age}_j}{\sum_{j} L_{{\rm bol},j}}\\
		&\langle [M/H]\rangle = \frac{\sum_{j} L_{{\rm bol},j}\times[M/H]_j}{\sum_{j} L_{{\rm bol},j}}.
	\end{align}
\end{subequations}

In \autoref{fig:mass_size} I show the distribution of ages, metallicities and the \citet{Sersic1968} index $n_{\rm Ser}$ on the $(M_{\rm JAM},R_{\rm e}^{\rm maj})$ plane, where the dynamical mass $M_{\rm JAM}$ closely approximates the total stellar masses $M_\ast$, and the half-light radius $R_{\rm e}^{\rm maj}$ is the semi-major axis of the isophote containing half of the total light of the \citet{Sersic1968} fitted to the galaxy photometry. Both $R_{\rm e}^{\rm maj}$ and $n_{\rm Ser}$ are tabulated in the LEGA-C catalogue \citep{vanderWel2021}. See \citet{Cappellari2013p15} for a discussion of why dynamical mass approximates the total stellar mass. 

I \textsc{loess} smoothed the measured values in all coloured plots of this paper. The resulting \textsc{loess}-smoothed values represent the two-dimensional equivalent of the average values that are routinely shown in one-dimensional plots. However, in two-dimension one cannot show the scatter together with the average values. To visualize the scatter, which is significant and not random, I use a projection along the $\sigma_*$ axes later.

This figure shows the well-known fact, in the nearby Universe, that both ages and metallicities approximately follow lines of constant stellar velocity dispersion $\sigma_\ast$, or equivalently lines where $M_\ast\propto R_{\rm e}^{\rm maj}$ (compare this figure with MaNGA results in fig.~4 of \citealt{Li2018} or \citealt{Lu2023}). I also show to guide the eye the local ``zone of avoidance'' at high densities \citep[eq.~4]{Cappellari2013p20}, which I scaled down by a factor $1.6\times$ in $R_{\rm e}^{\rm maj}$, roughly consistent with the general trends of decreasing galaxy sizes with redshift \citep[e.g.][]{vanderWel2014}. The Sersic index $n_{\rm Ser}$ also approximately follows the distribution of ages and metallicity, in the sense that passive galaxies tend to have $\ln n_{\rm Ser}\ga0.4$ or $n_{\rm Ser}\ga2.5$ (red colour in the right panel of \autoref{fig:mass_size}). This $n_{\rm Ser}$ value is the one sometimes adopted to separate early-type from late-type galaxies \citep[e.g.][]{Bell2003,Shen2003}. In the local Universe, below the stellar mass $\lg (M_\ast/\msun)\la11.5$ the trend of $n_{\rm Ser}$ is due to a sequence of increasing bulge fraction, while above $\lg (M_\ast/\msun)\ga11.5$ is the region of slow rotators with cores \citep[see review by][fig.~23]{Cappellari2016}.

This result has a rather long history, both locally and at $z\sim1$ \citep{Chauke2018,Chauke2019,Beverage2021,Barone2022,Hamadouche2022,Tacchella2022} but had not been seen so cleanly at this redshift before LEGA-C. 
For nearby galaxies, \citet{Kauffmann2003mass} clearly noted that galaxy population correlates better with mass surface density $\Sigma$ than with $M_\ast$. It was later observed that $\Sigma$, or even better the virial predictor $\sigma_{\rm vir}\propto M_\ast/R_{\rm e}$ of the stellar velocity dispersion, inferred from photometry alone, remains a better predictor of galaxy colours out to $z\approx3$ \citep{Franx2008,Bell2012}. However, it was still unclear at that time how accurately the photometric estimates were able to predict the actual stellar masses and the velocity dispersion of the stars. To address this issue I used masses from dynamical models, and $\sigma_\ast$ from good quality integral-field stellar kinematics, rather than photometric estimates. In \citet{Cappellari2011dur} I clearly concluded that ``$\sigma_\ast$ (not $\Sigma_{\rm e}$ or $M_\ast$) is the best predictor of galaxy properties'' \citep[see also][]{Cappellari2013p20}. These early results were confirmed by several papers using larger samples and stellar kinematics of ever-increasing quality \citep[e.g.][]{Wake2012,McDermid2015,Scott2017,Li2018,Barone2018,Barone2020}. In parallel, \citet{Cheung2012} and \citet{Fang2013} introduced the use of central surface density $\Sigma_1$ from photometry, within a fixed radius of 1 kpc, to predict quenching. A review is given in \citet[see fig.~22]{Cappellari2016}. 

Given that in \autoref{fig:mass_size} the main stellar population trends follow $\sigma_\ast$, in \autoref{fig:sig_age_metal} I show how the luminosity-weighted ages and metallicity depend on $\sigma_\ast$ in the LEGA-C sample. The trends resemble quite closely the local results from the best integral-field spectroscopy from both SAMI \citep{Scott2017} and MaNGA \citep{Li2018}.  However, the top panels of \autoref{fig:sig_age_metal} additionally illustrate the clear dependency between age and $[M/H]$ at fixed $\sigma_\ast$: the population of old galaxies at large $\sigma_\ast$ is characterized by a larger metallicity than their younger counterpart at the same $\sigma_\ast$. Very clear is the bend in the $(\sigma_\ast,{\rm Age})$ distribution around $\lg(\sigma_\ast/\kms)\approx2.3$ \citep[also see][]{Chauke2018}. The results are very consistent between both the \textsc{fsps} and \textsc{galaxev} SPS models. It is reassuring to see that the ridge of the age distribution in the top panels converges towards the age of the Universe at that redshift (grey horizontal band), while being slightly younger for \textsc{galaxev} vs \textsc{fsps}. I also run models where I restricted the age of each galaxy to the Universe's age at its redshift, as generally done for local studies. All results were qualitatively similar, except for the obvious truncation and corresponding clustering of the Ages values at the maximum Universe ${\rm Age}\approx6.6$ Gyr at $z\approx0.8$, which is indicated by a grey band in \autoref{fig:sig_age_metal}. The \textsc{bpass} results are qualitatively in agreement but show substantial quantitative differences, especially in the  $\sigma_\ast-[M/H]$ trend. Overall, this figure confirms the quality and consistency of these global results compared to local surveys.

\autoref{fig:dust} shows the distribution of the two dust attenuation parameters $A_V$ and $\delta$  (\autoref{sec:dust}) coloured by mean stellar age. One can see that at every UV slope $\delta$ the youngest galaxies have the strongest attenuation, except for the largest $\delta$. Moreover, the largest attenuations in galaxies are only observed at large $\delta$, close to the \citet{Calzetti2000} slope $\delta=0$. Note, however, that there is a degeneracy between attenuation and continuum normalization near the upper limit of $\delta$. Results are extremely consistent for the \textsc{fsps} and \textsc{galaxev} SPS models, but again the \textsc{bpass} results look quite different, although they all qualitatively agree.

\begin{figure*}
	\includegraphics[width=\textwidth]{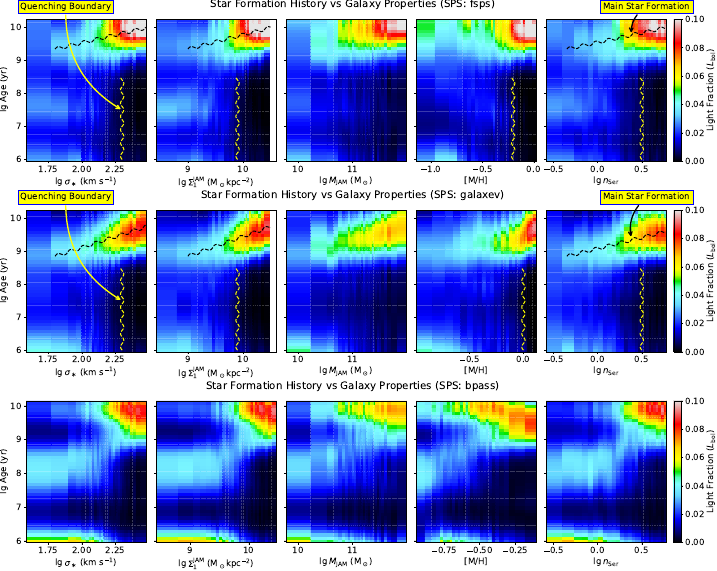}
	\caption{Star formation history (SFH) vs galaxy properties. In all panels, the colours represent the SFH recovered with \ppxf\ (with \texttt{regul=100}) parametrized by the bolometric light $L_{\rm bol}$ fraction contributed by populations of different ages. The SFHs are shown as a function of key galaxy parameters: (1) the stellar velocity dispersion $\sigma_\ast$; (2) the dynamical mass $M_{\rm JAM}$, which well approximates $M_\ast$; (3) the dynamically-determined average density $\Sigma_1^{\rm JAM}$ inside a circle of radius 1 kpc centred on the galaxy; (4) the luminosity weighted metallicity $\langle [M/H] \rangle$ and (5) the exponent $n_{\rm Ser}$ of a Sersic profile fitted to the galaxy photometry (see text for definitions). Average values are computed for equal bins in $\sigma_\ast$ (or the other parameters) of 100 galaxies each. Galaxies with largest $\sigma_\ast$, $M_\ast$, $\langle [M/H] \rangle$ or $n_{\rm Ser}$ on average experienced their main star formation event long ago, but the typical age for the bulk of their star formation increases with $\sigma_\ast$ (or the other parameters) as indicated by the slanted black dashed wavy lines in the left and right panels. For lower $\sigma_\ast$ galaxies can form the stars at any time until the present. The plots show a beautifully clear and sharp quenching boundary at $\lg(\sigma_\ast/\kms)\approx2.3$, or $\lg(\Sigma_1^{\rm JAM}/\mathrm{\msun kpc^{-2}})\approx9.9$, or $\langle [M/H] \rangle\approx-0.2$ for \textsc{fsps} and $\langle [M/H] \rangle\approx0.0$ for \textsc{galaxev}, or $\lg n_{\rm Ser}\approx0.5$, as indicated by the vertical yellow dashed wavy lines. There is no sharp boundary as a function of galaxy mass, but the transition is gradual and roughly happens around $\lg (M_\ast/\msun)\approx11.5$.  Note the generally good agreement between the results from the \textsc{fsps} and \textsc{galaxev} SPS. The results using the \textsc{bpass} models look problematic, with spurious structures at specific ages, which are most likely artefacts of the models.
		\label{fig:sfh}}
\end{figure*}

\begin{figure*}
	\includegraphics[width=\textwidth]{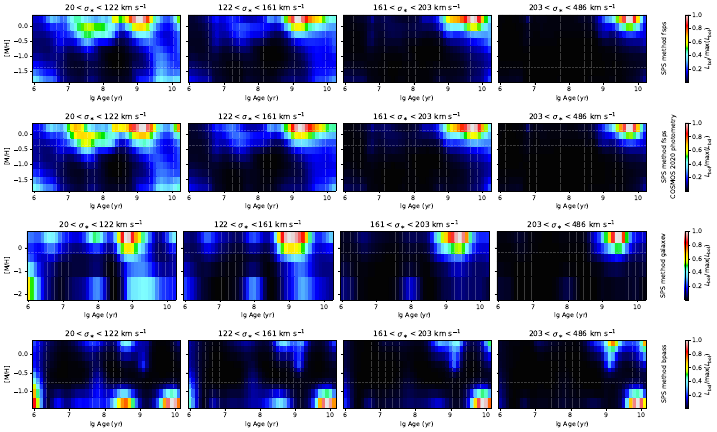}
	\caption{Joint star-formation history (SFH) and metallicity distributions. Each panel shows the average distribution of weights recovered with \ppxf\ (with \texttt{regul=10}) for 4 bins containing 800 LEGA-C galaxies sorted by their stellar velocity dispersion $\sigma_\ast$ (as indicated in the plot titles). The weights represent the bolometric luminosity $L_{\rm bol}$ contributed by populations of different ages and metallicities. The top row shows the results obtained with the \textsc{fsps} SPS models. The second row still uses \textsc{fsps} but adopts the COSMOS2020 \citep{Weaver2022} instead of my default UltraVISTA photometric catalogue \citep{Muzzin2013}. The third row is derived using the \textsc{galaxev} SPS models and the bottom one with the \textsc{bpass} models. One can see that the \textsc{fsps} and \textsc{galaxev} SPS provide qualitatively consistent results even for some of the main ``blobs'' in the distributions. Using an alternative photometric catalogue has virtually no effect on the results. At large $\sigma_\ast$ galaxies on average quenched long ago and their stars have high metallicity. At progressively lower $\sigma_\ast$ the age of the bulk of the star formation decreases, while still being dominated by high metallicity stars. However, the population is polluted by fresh accretion events of lower metallicity and a range of accretion times. As previously noted, the results using \textsc{bpass} SPS models are significantly different and should not be trusted, without further analysis.
		\label{fig:sfh_age_metal}}
\end{figure*}

\subsection{Non-parametric star formation histories}

\autoref{fig:sfh} shows the non-parametric star formation history of the galaxies in the LEGA-C sample as a function of key galaxy parameters. For this plot I sorted the quantity of interest (e.g. $\sigma_\ast$) and constructed 30 bins in that quantity, each containing the same number of about 100 galaxies, in such a way that different bins have the same level of shot noise. I show the dependency of the SFH, parametrized as discussed by the light $L_{\rm bol}$ contributed in the spectrum by stellar populations of different ages, as a function of the following parameters:
\begin{enumerate}
	\item \textbf{Stellar velocity dispersion:} the plots show a clear trend of SFH with $\sigma_\ast$ as expected from the trends between  $\sigma_\ast$ and age. What is new is the striking sharpness of the boundary between a regime $\lg(\sigma_\ast/\kms)\ga2.3$ (or $\sigma_\ast\ga200$ \kms), above which the spectra are dominated by a population nearly as old as the Universe at that redshift, without evidence for subsequent star formation events, and below which suddenly galaxies have star formation at any time until the present time. Both the \textsc{fsps} and \textsc{galaxev} SPS models indicate that galaxies still form the bulk of their stars at old times, but this age increases with $\sigma_\ast$ by roughly a factor between 6--10 for a variation in $\sigma_\ast$ by a factor of 10. In the case of the \textsc{galaxev} models, the SFH indicate ongoing star formation at the lowest $\sigma_\ast$ bins, while this is less so for the \textsc{fsps} models.
	
	\item \textbf{Density within 1 kpc:} this panel shows the same trend as the previous one, but with different units. This is because $\Sigma_1^{\rm JAM}$ is closely related to $\sigma_*$ \citep[see][]{Fang2013}, especially when using JAM dynamical masses instead of stellar population masses.
	
	\item \textbf{Galaxy mass:} contrary to the dependency of SFH with $\sigma_\ast$, there is no sharp transition as a function of stellar mass, but rather a gradual trend. Only galaxies more massive than $\lg (M_\ast/\msun)\ga11.5$ are characterized by a single event of star formation at old times. 
	
	\item \textbf{Galaxy metallicity:} this panel shows that $[M/H]$ is as good as $\sigma_\ast$ at predicting the boundary between the region of fully quenched galaxies and those that can have multiple star formation events. Here it happens at $[M/H]\approx-0.2$ for the \textsc{fsps} and $[M/H]\approx0.0$ for the \textsc{galaxev}, which are systematically shifted to larger values of metallicity.
	
	\item \textbf{Sersic index}: This panel show the SFH as a function of the \citet{Sersic1968} exponent $n_{\rm Ser}$. The boundary between fully quenched galaxies and galaxies that can have multiple star formation events happens here at $\lg n_{\rm Ser}\approx0.5$ or $n_{\rm Ser}\approx3.2$. Remarkably this boundary is here nearly as clean as that with $\sigma_\ast$.
\end{enumerate}
In all panels, the results using the \textsc{bpass} SPS models are again quite different from the other two. They show significant structure at specific ages and a less clear quenching boundary. The structure seen using the \textsc{bpass} models is likely an artefact of the SPS rather than a real conspiracy in the star formation events.

The trends for different galaxy parameters, and the overall consistency between the four panels, for both the \textsc{fsps} and \textsc{galaxev} SPS models, can be understood by looking at \autoref{fig:mass_size} and noting that there is a region $\lg (M_\ast/\msun)\ga11.5$ above which all galaxies are quenched, high metallicity and have large Sersic index. Below that mass galaxies follow a trend of increasing bulge fraction, which increases  $\sigma_\ast$, metallicity and makes galaxies more likely to quench. These results parallel those which have been extensively reported for local galaxies \citep[see review by][]{Cappellari2016}. What is new here is the clarity and sharpness of the empirical evidence of the boundary to quenching and the fact that this can be detected so well at a time when the Universe was half of its current age. 

A rapid cessation of star formation for galaxies above a given critical value of $\sigma_\ast$, or of some other estimate of the central stellar density, varying with $z$, has often been invoked to explain the evolution of galaxy parameters over time \citep[e.g.][]{vanDokkum2015}. An excellent review of the empirical evidence and models of a quenching boundary in galaxies is given in \citet[sec.~1]{Chen2020}. The physical mechanism for quenching is still under debate. Very briefly, one can group the main proposed theories into three broad classes: (i) ``halo quenching'', where the gas gets shock-heated when falling into the gravitational potential of massive dark halos \citep[e.g.][]{Dekel2006}; (ii) ``active galactic nucleus (AGN) feedback'', where a jet from the supermassive black hole either ejects the gas from its host galaxy \citep[e.g.][]{Silk1998} or prevent it from infalling \citep[e.g.][]{Bower2006,Croton2006}. See review by \citet{Somerville2015}. The panels in \autoref{fig:sfh} provide a beautiful empirical confirmation of the theoretical assumptions that are made in many of those models.

\subsection{Non-parametric joint SFH and metallicity distributions}

\autoref{fig:sfh_age_metal} presents the non-parametric joint luminosity distribution of the age and metallicities of the stellar populations of galaxies in four different bins of $\sigma_\ast$. As in \autoref{fig:sfh}, also for this figure I sorted galaxies as a function of their $\sigma_\ast$ and constructed four groups, of about 800 galaxies each, to ensure all panels have the same level of shot noise. Like before, I compare all three SPS models (\textsc{fsps}, \textsc{galaxev} and \textsc{bpass}). In addition, In the second row of \autoref{fig:sfh_age_metal} I show the result when using the \textsc{fsps} model but adopting the photometric measurements from COSMOS2020 \citep{Weaver2022} instead of the UltraVISTA catalogue \citep{Muzzin2013}. The first and second rows are barely distinguishable and this shows that any possible difference in the photometric calibration has a completely insignificant effect on the results. The distribution from both \textsc{fsps} and \textsc{galaxev} is highly consistent, almost at the level of the individual ``blobs'', except for slightly older younger ages and higher metallicities for the \textsc{galaxev} vs the \textsc{fsps} models. 

The plots indicate that even galaxies with low $\sigma_\ast$ are still dominated by stars with high metallicity, but this is diluted by extra lower-metallicity populations acquired at different times. I should stress that the relatively smooth distribution in the maps are averages of many galaxies and should not be interpreted as the evolution of one individual galaxy, which is generally characterized by discrete star formation events. Moreover, not every feature of the maps is robust against variations in the data and SPS models.

The overall observed distribution could be interpreted in the context of the two-phases of galaxy formation \citep[e.g.][]{Oser2010}. According to this scenario, the formation of galaxies has a “two-phase” nature: a fast initial phase at $z \ga 2$ where “in situ” stars are created inside the galaxy from cold gas that falls in, and a longer phase since $z \la 3$ where “ex-situ” already-formed stars are mainly acquired. In this phase, large systems increase their mass and radius by absorbing smaller stellar systems that were formed very early ($z \ga 3$) outside of the central galaxy's virial radius, or by smooth gas accretion from cosmological filaments \citep[see][for a review]{Naab2017}. 
	
Specifically, the old, high-metallicity component observed in \autoref{fig:sfh_age_metal} could be interpreted as the relic of the in-situ formation, which was quickly metal enriched, while the lower metallicity would correspond to either acquired stars, previously formed in smaller stellar components, or to star formation due to accretion from low-metallicity cosmological filaments. The accreted component is only present below the critical ``quenching boundary'' of $\sigma_*\la200$ \kms. Below that boundary, accretion can continue throughout the galaxies evolution. The high-metallicity old peak is visible for all four subsets of $\sigma_*$, but its age decreases with $\sigma_*$. This age trend in the old-age peak is the same already pointed out in \autoref{fig:sfh}.

The LEGA-C spectra I analysed are not spatially resolved, but a similar analysis of spatially-resolved integral-field spectroscopic data for the MaNGA survey shows that, in low $\sigma_*$ galaxies in the nearby Universe, the oldest higher-metallicity component is associated to the galaxy bulge, while low-metallicity gas accretion happens in the disk \citep[e.g.][]{Lu2023}.

As expected, the \textsc{bpass} models show again quite different results, with a markedly different metallicity distribution. As commented earlier, the results from this model should be treated with caution as they are likely dominated by spurious unknown effects in the models.

A caveat on these results on the metallicity distribution, which also affects other similar results on metallicity determinations from galaxy stellar spectra, is that the signature of metallicity variations becomes weaker at younger ages, where the $S/N$ of the data also generally decreases. This can introduce possible systematic effects on metallicity trends. To exclude the effect of $S/N$, I verified that all results remain unchanged if I restrict the analysis to the 873 galaxies with $S/N>20$ and even, at coarser resolution, for the subset of 126 galaxies with $S/N>40$. It would still be valuable to compare the reported metallicity trends e.g. with those inferred from gas tracers from similar data.

\section{Summary}
\label{sec:summary}

In the first half of this paper, I described some modifications to the \ppxf\ method \citep{Cappellari2017}, which is used to extract the stellar and gas kinematics, as well as the stellar population of galaxies. First, I described a novel constrained least-squares optimization algorithm that \ppxf\ has been using for the past few years. Then I outlined the changes I made to \ppxf\ to be able to fit photometric data together with the usual full-spectrum fitting. I also described some other minor changes.

In the second half of the paper, I presented an application of \ppxf\ to the extraction of non-parametric star formation histories and metallicity distributions for a sample of 3200 galaxies at redshift $0.6<z<1$ with spectroscopy from the LEGA-C survey DR3 \citep{vanderWel2021}, and with 28-bands photometric measurements covering from the far ultraviolet (0.1 \micron) to the near-infrared (3 \micron) from either the UltraVISTA  \citep{Muzzin2013} or the COSMOS2020 catalogues \citep{Weaver2022}. I also constructed JAM dynamical models \citep{Cappellari2008,Cappellari2020} for all galaxies with measured stellar dispersion $\sigma_*$ and available Sersic profile fits to the photometry.

For this study, I used and compared three spectral population synthesis (SPS) methods satisfying some criteria of age and wavelength coverage. This led to my selection of the \textsc{fsps} \citep{Conroy2009,Conroy2010}, \textsc{galaxev} \citep{Bruzual2003} and \textsc{bpass} \citep{Stanway2018,Byrne2022} SPS methods.

I compared the dynamical masses from JAM against the stellar masses from the different stellar-population fitting methods. I found that \ppxf\ with photometry and spectra provides more accurate masses than the other methods with photometry alone, as one would have expected.

I found that \ppxf\ on these data reveals a striking difference between galaxies that are only consistent with a single star formation event from those that require multiple bursts of star formation.

I constructed scaling relations for the global stellar population parameters and found a remarkable similarity, but even clearer trends, between these results at $z\approx0.8$ and those from the latest spectroscopic surveys in the nearby Universe. This gives some confidence in the meaningfulness of the results and highlights the quality of the spectro-photometric data.

Finally, I explored the non-parametric star formation histories (SFH) and the joint SFH and metallicity $[M/H]$ distributions. I found that the data indicate, on average over many galaxies, a remarkably sharp quenching boundary for the cessation of star formation, at a stellar velocity dispersion $\lg(\sigma_\ast/\kms)\approx2.3$ ($\sigma_\ast\approx200$ \kms), or equivalently with average mass density within 1~kpc $\lg(\Sigma_1^{\rm JAM}/\mathrm{\msun kpc^{-2}})\ga9.9$ ($\Sigma_1^{\rm JAM}\ga7.9\times10^9\,\mathrm{\msun\ kpc^{-2}}$), or at metallicity $[M/H]\approx-0.1$ (with some variation dependent on the adopted SPS model) or at \citet{Sersic1968} index $\lg n_{\rm Ser}\approx0.5$ ($n_{\rm Ser}\approx3.2$). As expected, the transition is more gradual as a function of stellar mass. This abrupt quenching boundary has been invoked by several models of galaxy formation. These data provide one of the cleanest empirical evidence to date. 

The joint age-metallicity distribution appears to support the two-phase scenario of galaxy evolution by revealing the relic of an old quickly-formed high-metallicity component and, below the quenching boundary $\sigma_\ast\la200$, multiple events of lower-metallicity accretion.

This paper only scratches the surface of what can be done with this dataset and with similar ones that are being acquired at comparable and higher redshift. I have not explored e.g. obvious dependencies between SFH and stellar kinematics or environment \citep[e.g.][]{Cole2020,Sobral2022}. Comparisons with galaxy formation models should be performed in the space of observable rather than using stellar masses which are empirically more uncertain. A similar analysis at higher redshift can reveal the onset and variation of the quenching boundary, which is a key but still quite uncertain parameter in galaxy formation models. James Webb Space Telescope (JWST) data are ideal to extend this kind of study to higher redshift.

\section*{Acknowledgements}

I am grateful to the referee for an expert and very useful report.
Based on observations made with ESO Telescopes at the La
Silla Paranal Observatory under program IDs 194-A.2005 and
1100.A-0949 (The LEGA-C Public Spectroscopy Survey).

\section*{Data Availability}

The LEGA-C DR3 spectra and catalogue are available   \href{http://archive.eso.org/cms/eso-archive-news/Third-and-final-release-of-the-Large-Early-Galaxy-Census-LEGA-C-Spectroscopic-Public-Survey-published.html}{HERE}, the UltraVISTA photometric catalogue  \href{https://local.strw.leidenuniv.nl/galaxyevolution/ULTRAVISTA/Ultravista/UltraVISTA_Catalog_Home.html}{HERE}, the COSMOS2020 catalogue from \url{https://cosmos2020.calet.org/}, the \ppxf\ software from \url{https://pypi.org/project/ppxf/}, the JAM software from \url{https://pypi.org/project/jampy/}, the \textsc{MgeFit} software from \url{https://pypi.org/project/mgefit/} and the \textsc{LtsFit} software from \url{https://pypi.org/project/ltsfit/}.

\bsp    % typesetting comment
\label{lastpage}

\end{document}